\pdfoutput=1
\documentclass[pra,twocolumn,showpacs,amsmath,amssymb,floatfix]{revtex4-1}

\usepackage{graphicx}%Include figure files
\usepackage{dcolumn}%Align table columns on decimal point
\usepackage{bm}% bold math
\usepackage[colorlinks=true, linkcolor=blue, urlcolor=blue,
citecolor=blue]{hyperref}
\usepackage{mathrsfs}

\begin{document}

\title{Spin-orbit-proximitized ferromagnetic metal by monolayer transition metal dichalcogenide: Atlas of spectral functions, spin textures and spin-orbit torques in Co/MoSe$_2$, Co/WSe$_2$ and Co/TaSe$_2$ heterostructures}

\author{Kapildeb Dolui}
\affiliation{Department of Physics and Astronomy, University of Delaware, Newark, DE 19716-2570, USA}
\author{Branislav K. Nikoli\'{c}}
\email{bnikolic@udel.edu}
\affiliation{Department of Physics and Astronomy, University of Delaware, Newark, DE 19716-2570, USA}

\begin{abstract}
The bilayer heterostructures composed of an ultrathin ferromagnetic metal (FM) and a material hosting strong spin-orbit (SO) coupling are principal resource for SO torque and spin-to-charge conversion nonequilibrium effects in spintronics. The key quantity in theoretical description of these effects is {\em  nonequilibrium spin density}, which can appear on any monolayer of the heterostructure through which the current is flowing while the monolayer bands are affected by the native or proximity induced SO coupling. Here we demonstrate how hybridization of wavefunctions of Co layer and a monolayer of transition metal dichalcogenides (TMDs)---such as semiconducting MoSe$_2$ and WSe$_2$ or metallic TaSe$_2$---can lead to {\em dramatic transmutation} of electronic and spin structure of Co within some distance away from its interface with TMD,  when compared to the bulk of Co or its surface in contact with vacuum. This is due to proximity induced SO splitting of Co bands encoded in the spectral functions and spin textures on its monolayers, which we obtain using noncollinear density functional theory (ncDFT) combined with equilibrium Green function (GF) calculations. In fact, SO splitting is present due to structural inversion asymmetry of the bilayer even if SO coupling within TMD monolayer is artificially switched off in ncDFT calculations, but switching it on makes the effects associated with proximity SO coupling within Co layer about five  times larger. Injecting spin-unpolarized charge current through SO-proximitized monolayers of Co generates nonequilibrium spin density over them, so that its cross product with the magnetization of Co determines SO torque. The SO torque computed via  first-principles quantum transport methodology, which combines ncDFT with nonequilibrium GF calculations, can be used as the screening parameter to identify optimal combination of materials and their interfaces for applications in spintronics. In particular, we  identify heterostructure two-monolayer-Co/monolayer-WSe$_2$  as {\em the most optimal}.
\end{abstract}

%\pacs{72.25.Dc, 75.70.Tj, 71.15.Mb, 72.10.Bg}
\maketitle

\section{Introduction} \label{sec:intro}

The study of proximity effects, such as superconducting and magnetic, within bulk normal materials has a long history~\cite{Kircher1968,Hauser1969}. It has been  rekindled with the advent of new materials and/or regimes of quantum coherence. For example, mesoscopic regime of normal materials~\cite{Altland2000,Gueron1996} at low temperatures host large-scale phase-coherent single electron states which can couple to macroscopically coherent many-body state on the superconducting side of  the interface; both superconducting and magnetic proximity effects have been intensely explored~\cite{Zutic2019} as soon as two-dimensional (2D) materials and their van der Waals  heterostructures have been discovered. In the superconducting proximity effect~\cite{Altland2000,Gueron1996,Freericks2001,Freericks2002,Nikolic2002b}, Cooper pairs leak from the superconductor to the normal material side, where the density of Cooper pairs decreases exponentially from superconductor/normal-material interface. Such  creation of correlated electron-hole pairs and the corresponding change in the local properties of the normal material have been explored in metals~\cite{Kircher1968,Gueron1996}, ferromagnets~\cite{Buzdin2005}, 2D~\cite{Komatsu2012} and topological materials~\cite{Li2018a} brought into the contact with a superconductor. The {\em direct} effect is also accompanied by the {\em  inverse} superconducting proximity effect~\cite{Nikolic2002b,Sillanpaa2001} where the order parameter is depleted  and electronic density of states is induced inside the superconducting  energy gap within some length of the superconductor side of the interface. 

\begin{figure}
	\includegraphics[width=8.0cm,clip=true]{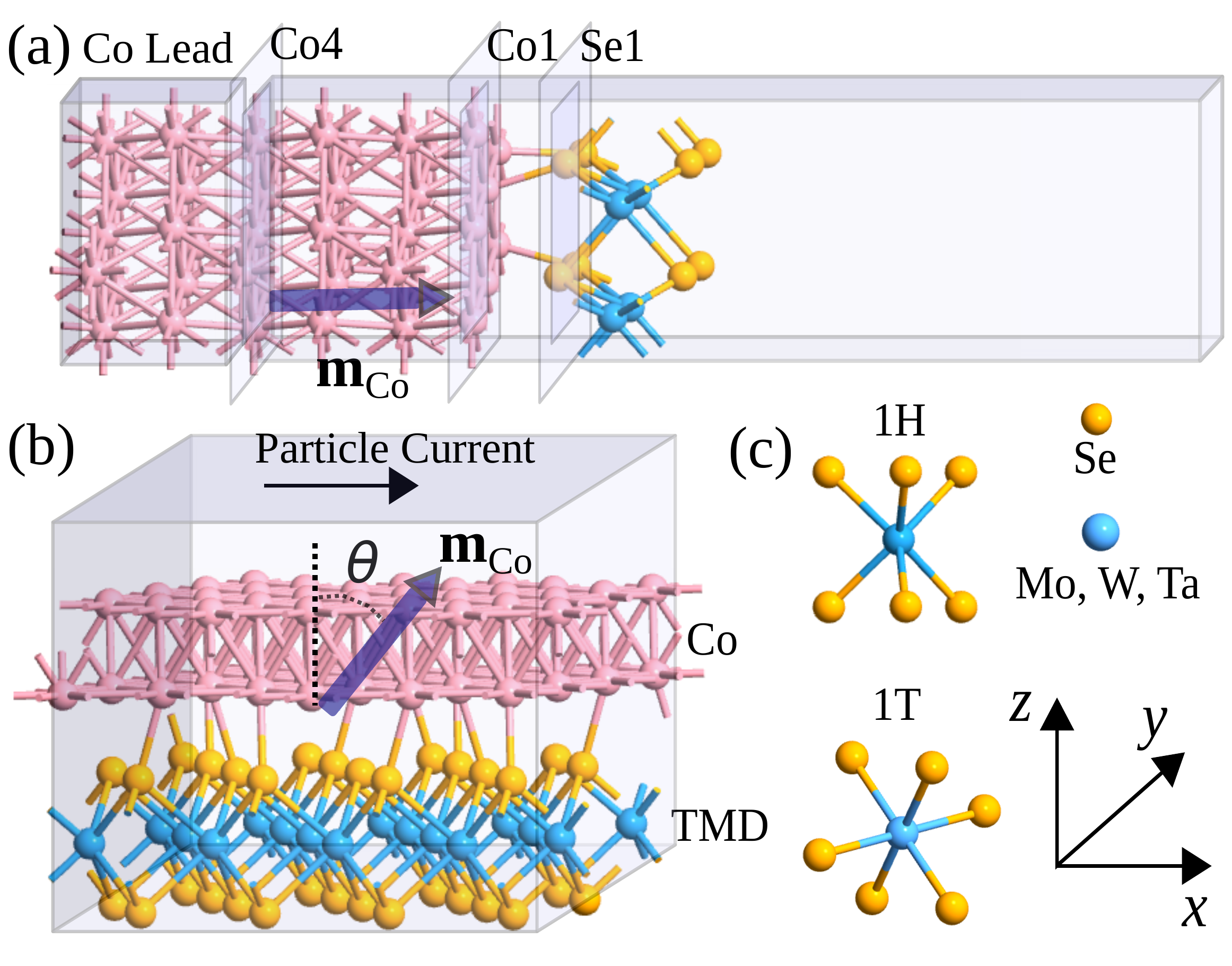}
	\caption{(a) Schematic view of heterostructure semi-infinite Co(0001)/monolayer-TMD---where TMD=MoSe$_2$, WSe$_2$, or TaSe$_2$---which is employed to calculate spectral function using Eq.~\eqref{eq:spectral} and the corresponding spin textures. These quantities are computed on planes passing through Se and Co atomic layer at the interface, denoted by Se1 and Co1, respectively, as well as on the 4th ML of Co away from the interface which is  denoted by Co4. (b) Lateral heterostructure 2ML-Co/monolayer-TMD, with unpolarized current injected parallel to its interface using small bias voltage $V_b$ along the $x$-axis in the linear-response regime. The heterostructure in (a) is assumed to be infinite in the $yz$ plane, while the heterostructure in (b) is infinite in the $xy$ plane. (c) Schematic view of 1H- and 1T-polytype structure of TMDs, in which transition metal atom has trigonal prismatic and octahedral coordination with Se atoms, respectively. 1T structure preserves the spatial inversion symmetry, where 1H structure does not.}
	\label{fig:fig1}
\end{figure}

\begin{figure}
	%\center
	\includegraphics[width= \linewidth,clip=true]{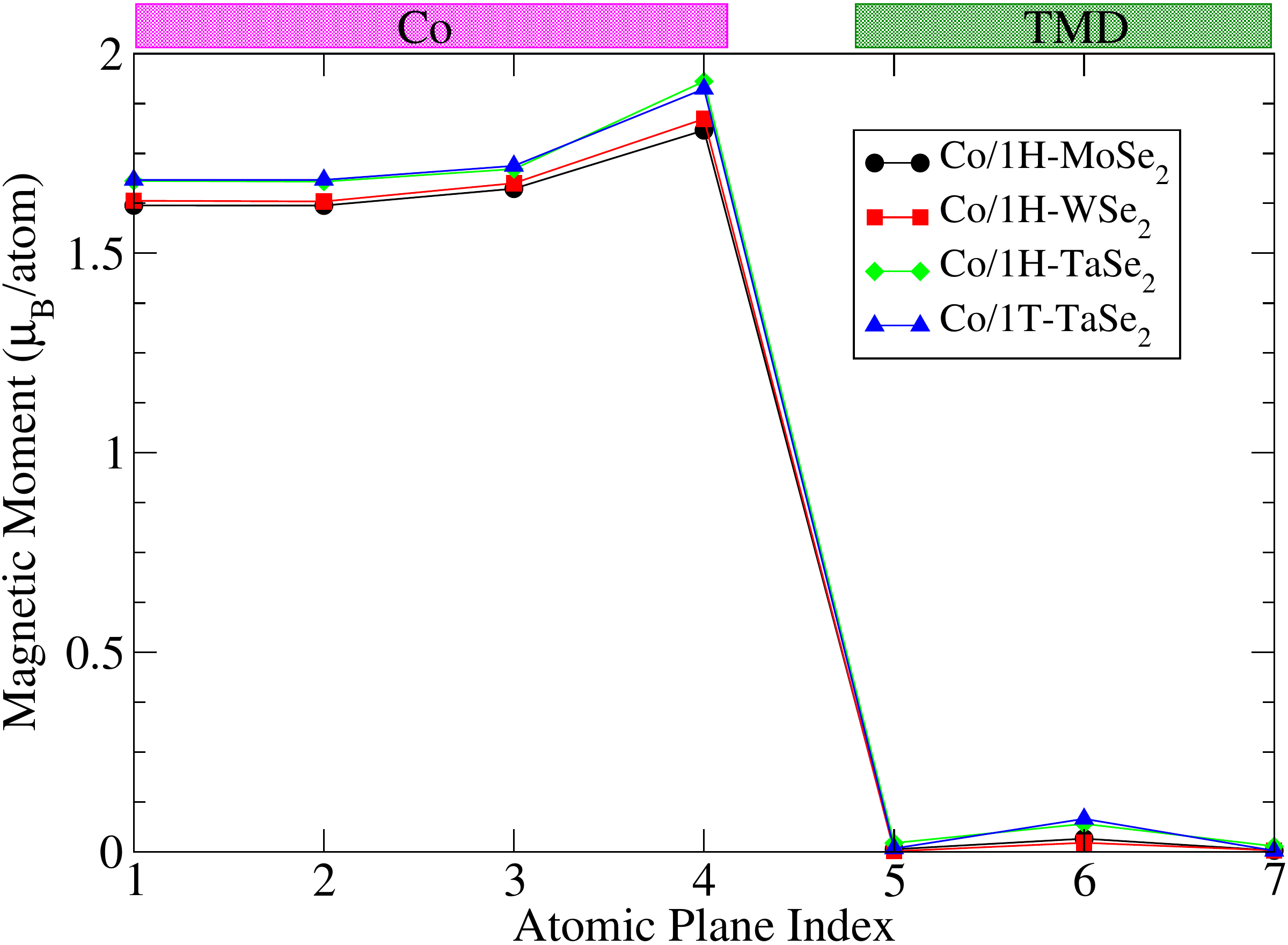}
	\caption{Spatial profile of magnetic moments, averaged over all atoms within each atomic plane, for semi-infinite-FM/monolayer-TMD heterostructures illustrated in Fig.~\ref{fig:fig1}(a): (a) Co/1H-MoSe$_2$ (black circles); (b) Co/1H-WSe$_2$ (red squares); (c) Co/1H-TaSe$_2$ (green diamonds); and (d) Co/1T-TaSe$_2$ (blue triangles). Magenta and green solid boxes on the top of the panel show the spatial extent of the Co and TMD layers, respectively, as a guide to the eye.}
	\label{fig:fig2}
\end{figure}

The analogous {\em direct} magnetic proximity effect emerges when ferromagnetic metal (FM) induces nonzero local magnetization in the adjacent non-magnetic material which decays exponentially away from the interface~\cite{Hauser1969,Lim2013,Zhu2018,Peterson2018,Belashchenko2019}. In FM/TMD heterostructures in the focus of this study, where FM is Co and TMD is monolayer (ML) of transition metal dichalcogenide (TMD), as illustrated in Fig.~\ref{fig:fig1}, the nonzero magnetic moment in the setup of Fig.~\ref{fig:fig1}(a) appears in Fig.~\ref{fig:fig2} on the middle atomic plane of TMD monolayer. The layered structure of TMDs of the type MX$_2$ in Fig.~\ref{fig:fig1}---where M=Mo,W,Ta and X=Se---is formed by graphene-like hexagonal arrangement of M and X atoms bonded together to give X-M-X sandwich of three atomic planes. Within the sandwich, each M atom is covalently bonded to six X atoms, whereas each X atom is connected to three M atoms.  For all TMDs considered in Fig.~\ref{fig:fig2}, the bands around the Fermi level are mainly due to $d$-orbitals of the respective transition metal. Consequently, the Co layer  induces an one order magnitude larger magnetic moment ($\sim$ 0.03~$\mu_B$~per metal atom) at the transition metal atomic plane M, in comparison to that on the X=Se atomic planes.

In contrast to the inverse superconducting proximity effect, the inverse magnetic proximity effect can manifest either as enhancement or suppression of magnetic order on the FM side of the interface. For example, {\em enhanced} magnetic moments appear on the first ML of Co (denoted as atomic plane 4) in Fig.~\ref{fig:fig2} which is in direct contact with monolayer TMD.  Conversely, {\em reduced}  magnetic moments in Co layer near the interface with 
heavy metals Pt or Ta are found in first-principles calculations in Fig.~4 of Ref.~\cite{Dolui2017}. 

Recently the ``proximity effect'' terminology has been ported to describe how one material can be dramatically transformed by acquiring properties of 
its neighbor~\cite{Zutic2019}, even though the property acquired is not traditionally studied proximity superconductivity or magnetism. For example, normal metal~\cite{Shoman2015,Spataru2014} or FM layer~\cite{Marmolejo-Tejada2017,Zhang2016,Hsu2017} in contact with a topological insulator (TI) can acquire spin-momentum locking in metallic band structure, as one of the salient features of TI~\cite{Bansil2016}, even though the Dirac cone energy-momentum dispersion on  the TI side of their interface is heavily distorted by hybridization~\cite{Marmolejo-Tejada2017,Hsu2017}. The long sought room-temperature noncollinear magnetic textures called skyrmions, with  nontrivial topology in real space, have been achieved experimentally by moving away from bulk materials~\cite{Nagaosa2013} to interfaces of ultrathin FM layers (composed of $\le 3$ MLs) and heavy metals~\cite{Boulle2016,Woo2016,Soumyanarayanan2017} which impart strong Dzyaloshinskii-Moriya interaction between magnetic moments on the FM side~\cite{Soumyanarayanan2016,Simon2018}. Once the inversion symmetry is broken due to the surface of FM or metals (such as Au, Ag or Cu~\cite{Tamai2013}) with vacuum, the Rashba type of spin-orbit (SO) coupling emerges in their Shockley surface state and it can penetrates over few MLs into the bulk~\cite{Marmolejo-Tejada2017}. Such interfacial SO coupling remains present around the interface, even if the surface is covered by normal metal without strong SO coupling effects, as exemplified by inversion asymmetric Cu/Co/Cu heterostructure~\cite{Dolui2017} or by Co layer which remains SO-proximitized [Fig.~\ref{fig:fig6}]  despite SO coupling being artificially switched off on the TMD side of Co/TMD interface in Fig.~\ref{fig:fig1}(b).

As regards the traditional ``proximity effect'' terminology, which suggests  fast decay of the proximity-acquired quantity away from the interface in analogy with superconducting and magnetic proximity effects, it is instructive to clarify different possibilities using examples of SO~\cite{Gmitra2016,Zollner2019b,Frank2016,Island2019} or magnetic proximity effects~\cite{Dahal2014,Lazic2016,Hallal2017} in (monolayer or multilayer) graphene~\cite{Zollner2019b,Frank2016,Island2019} and 2D materials~\cite{Dolui2020}. For example, when evanescent wavefunction from semimetallic graphene penetrates into a ferromagnetic insulator (FI) where there are no states at the Fermi level, it acquires exchange splitting from the native ferromagnetism of the FI layer~\cite{Hallal2017}. Conversely, propagating wavefunction on the metallic surface of TI (or even normal metal like Cu~\cite{Frank2016}) or FM hybridizes with  wavefunction of graphene, thereby directly SO-splitting~\cite{Zollner2019b} or spin-polarizing~\cite{Zollner2019b,Dahal2014,Lazic2016} electronic structure of graphene, respectively. The SO coupling  can also be introduced by evanescent wavefunctions, as exemplified by metallic TaSe$_2$ in contact with insulating antiferromagnetic bilayer CrI$_3$ in which case SO proximity effect is exponentially decaying and it appears only on the first ML of CrI$_3$ in direct contact with TaSe$_2$~\cite{Dolui2020}. Such SO coupling ``injected'' by the proximity effect is also quite distinct from the native  SO coupling of CrI$_3$ which is responsible for its magnetocrystalline anisotropy which stabilizes low-dimensional magnetic ordering at low temperatures. Thus, when proximity effect is mediated by evanescent wavefunctions, it decays exponentially fast away from the interface akin to traditional superconducting and magnetic proximity effects. On the other hand, when it is mediated by hybridization of propagating wavefunctions, the acquired property   can propagate over long distances away from the interface (at least in clean systems). For heterostructures in Fig.~\ref{fig:fig1}(a), this is exemplified by the presence of proximity SO coupling-induced in-plane spin textures  in Fig.~\ref{fig:fig3} on the 1st ML Co1 directly at the interface, as well as on the 4th ML Co4 away from the interface. Thus, an ultrathin Co layer, such as 2 MLs of Co in Fig.~\ref{fig:fig1}(b) or up to 8 MLs employed in Fig.~\ref{fig:fig7}, will be fully SO proximitized over their whole volume. On the other hand, the SO proximity effect will slowly (but not exponentially) decay~\cite{Marmolejo-Tejada2017} into the bulk of semi-infinite layer in Fig.~\ref{fig:fig1}(a).

The SO-proximitized FM have emerged as a major resource for spintronics, enabling effects like SO torque~\cite{Manchon2019,Ramaswamy2018}, spin-to-charge conversion~\cite{Han2018} and skyrmion generation~\cite{Soumyanarayanan2016}. In the SO torque phenomenon, injecting unpolarized charge current parallel to the interface FM/SO-coupled-material induces current-driven (CD) nonequilibrium spin density $\mathbf{S}_\mathrm{CD}$ which then drives the dynamics of the magnetization of FM layer via the SO torque $\propto \mathbf{S}_\mathrm{CD}  \times \mathbf{m}$~\cite{Manchon2019}, where $\mathbf{m}$ is the unit vector along the magnetization direction. Reciprocally, if nonequilibrium spin density is injected into the bilayer, such as by pumping of spin current due to precessing magnetization driven by microwaves~\cite{Tserkovnyak2005,Chen2009,Mahfouzi2012,Dolui2019}, spin-to-charge conversion is initiated  leading to charge current generation even in the absence of any bias voltage~\cite{Sanchez2013,Mahfouzi2014a,Shen2014}. Although early theoretical studies of SO torque and spin-to-charge conversion assume simplistic model Hamiltonians of the interface~\cite{Shen2014,Manchon2008,Lee2015,Ndiaye2017}, it has been realized that three-dimensional (3D) geometry of transport is crucial to capture all relevant effects~\cite{Mahfouzi2014a,Kim2017,Amin2018,Ghosh2018}. 

The simplistic tight-binding models of FM/SO-coupled-material bilayers can already capture some aspects of SO proximity effect~\cite{Ghosh2018}, but first-principles calculations are required to accurately describe charge transfer, surface relaxation and band bending between the two materials~\cite{Marmolejo-Tejada2017,Spataru2014,Zhang2016,Hsu2017}. In this study, we delineate how properties of conventional room temperature FM, such as semi-infinite Co layer in the region near the interface in Fig.~\ref{fig:fig1}(a) or whole volume of ultrathin (of thickness $\sim 1$ nm) Co layer in Fig.~\ref{fig:fig1}(b), can {\em dramatically} change due to just a single ML of adjacent TMD. Describing effects like complex spin textures  [Fig.~\ref{fig:fig3}] within Co in equilibrium due to proximity SO coupling yields accurate first-principles Hamiltonian as an input for quantum transport calculations of CD nonequilibrium spin density $\mathbf{S}_\mathrm{CD}$ [Figs.~\ref{fig:fig4} and ~\ref{fig:fig5}] on MLs of Co that host such textures. This also allows us to compute the corresponding SO torque $\propto \mathbf{S}_\mathrm{CD} \times \mathbf{m}_\mathrm{Co}$ [Figs.~\ref{fig:fig5}, ~\ref{fig:fig6} and ~\ref{fig:fig7}]. 

Using these quantities as descriptors for screening public databases of band structures of 2D materials~\cite{Haastrup2018,Zhou2019a} and their heterostructures~\cite{Andersen2015}, or for screening  ``mini database'' of spectral functions and spin textures [Fig.~\ref{fig:fig3}] created in this study for specific case of FM/TMD heterostructures, makes it possible to precisely  identify combinations that maximize features of relevance for spintronics applications. For example, for Co/TMD heterostructure and three different TMDs investigated, we identify Co/1H-WSe$_2$ as the optimal combination for maximum SO torque, as demonstrated in Fig.~\ref{fig:fig5}. The complex angular dependence of SO torque in  Co/1H-WSe$_2$ heterostructure is then investigated further in Fig.~\ref{fig:fig6} where we demonstrate that even with SO coupling is artificially switched off on the WSe$_2$ side, $\mathbf{S}_\mathrm{CD}$  and thereby driven SO torque are nonzero due to overall  broken inversion symmetry of the bilayer. Moreover, upon switching on the SO coupling in the first-principles calculations, SO torque is {\em enhanced} by a factor of five in Fig.~\ref{fig:fig6} and Table~\ref{tab:tab1}, thereby further highlighting the value of computational screening of heterostructures by combined first-principles and quantum transport calculations. 

The paper is organized as follows. In Sec.~\ref{sec:methods} we explain our first-principles quantum transport algorithm which combines nonequilibrium Green function (NEGF)~\cite{Stefanucci2013} with noncollinear density functional theory (ncDFT) calculations to  directly compute SO torque components that are even or odd in the magnetization of FM layer. Section~\ref{sec:spectral} shows the results for spectral functions and spin textures in equilibrium on both side of Co/TMD interface in Fig.~\ref{fig:fig1}(a). In Sec.~\ref{sec:sot}, we discuss CD nonequilibrium spin density and thereby driven SO torque in Co/TMD lateral heterostructure depicted in Fig.~\ref{fig:fig1}(b) where unpolarized charge current is injected parallel to the interface. We conclude in Sec.~\ref{sec:conclusions}.

\section{Models and Methods}\label{sec:methods}

\subsection{Spin-orbit torque from nonequilibrium density matrix expressed using NEGF}\label{sec:negf}

Both conventional spin-transfer torque, in spin valves or magnetic tunnel junctions containing two FM layers with noncollinear magnetizations~\cite{Wang2008b,Ellis2017,Nikolic2018}, and SO torque~\cite{Manchon2019}, in setups such as the one in Fig.~\ref{fig:fig1}(b) 
containing {\em only one} FM layer, can be described microscopically and independently of particular physical  mechanism~\cite{Belashchenko2020,Mahfouzi2018,Mahfouzi2020} in a {\em unified} fashion as  a consequence of local CD nonequilibrium spin density of conduction electrons $\mathbf{S}_{\rm CD}(\mathbf{r})$~\cite{Nikolic2018,Belashchenko2019}. The cross product of this quantity with the local magnetization $\mathbf{M}(\mathbf{r})$ in the case of simplistic model  Hamiltonians~\cite{Lee2015},  or with the exchange-correlation (XC) magnetic field $\mathbf{B}_\mathrm{XC}(\mathbf{r})$ of ncDFT~\cite{Capelle2001,Eich2013a} in the case of first-principles Hamiltonians~\cite{Nikolic2018,Ellis2017,Belashchenko2019,Belashchenko2020,Freimuth2014}, determines local spin torque $\mathbf{S}_\mathrm{CD}(\mathbf{r}) \times \mathbf{B}_\mathrm{XC}(\mathbf{r})$ at some point in space $\mathbf{r}$. Thus, the total torque is obtained by integrating over the FM layer receiving the torque
\begin{equation}\label{eq:sotrealspace}
\mathbf{T}_\mathrm{CD} = \int_\mathrm{FM} \! d^3r \, \mathbf{S}_\mathrm{CD}(\mathbf{r}) \times \mathbf{B}_\mathrm{XC}(\mathbf{r}).
\end{equation}
Here $\mathbf{S}_\mathrm{CD}(\mathbf{r})=\mathrm{Tr}\,[{\bm \rho}_\mathrm{CD} {\bm \sigma}]$ is computed by tracing 
the CD contribution to the nonequilibrium density matrix~\cite{Mahfouzi2013}
\begin{equation}\label{eq:rhocd}
{\bm \rho}_\mathrm{CD} = {\bm \rho}_\mathrm{neq} - {\bm \rho}_\mathrm{eq}
\end{equation}
with the vector of the Pauli matrices ${\bm \sigma} = (\hat{\sigma}_x, \hat{\sigma}_y,\hat{\sigma}_z)$, where ${\bm \rho}_\mathrm{eq}$ is the 
grand canonical density matrix of electrons in equilibrium~\cite{Stefanucci2013,Mahfouzi2013}. The lesser Green function (GF)
\begin{equation}\label{eq:lesser}
\mathbf{G}^{<}(E) = i \mathbf{G}(E)[f_\mathrm{L}(E)\boldsymbol{\Gamma}_\mathrm{L}(E)+f_\mathrm{R}(E){\bm \Gamma}_\mathrm{R}(E)]\mathbf{G}^{\dagger}(E),
\end{equation}
of NEGF formalism~\cite{Stefanucci2013} offers an efficient route to compute the nonequilibrium density matrix in the steady-state and elastic transport regime
\begin{equation}\label{eq:noneqrho}
\boldsymbol{\rho}_{\rm neq} = \frac{1}{2\pi i} \int_{-\infty}^{\infty}\!\! dE\, \mathbf{G}^{<}(E),
\end{equation}
for arbitrary periodic or nonperiodic device setup by splitting it into two semi-infinite leads and the active region. In the case of SO torque, one typically~\cite{Belashchenko2019,Belashchenko2020,Freimuth2014,Mahfouzi2018,Mahfouzi2020} considers periodic setup in the $xy$-plane, as exemplified by 
Co/TMD heterostructures in Fig.~\ref{fig:fig1}(b) whose semi-infinite leads and active region, of conveniently chosen length, are made of 
identical atoms. Here 
\begin{equation}\label{eq:retarded}
\mathbf{G} =  [E {\bm \Lambda} - \mathbf{H} -\boldsymbol{\Sigma}_\mathrm{L}(E,V_\mathrm{L})-\boldsymbol{\Sigma}_\mathrm{R}(E,V_\mathrm{R})]^{-1}
\end{equation}
is the retarded GF; $\mathbf{H}$ is the matrix representation of (model of first-principles) Hamiltonian of the active region in some basis of localized orbitals basis  $|\phi_i\rangle$; \mbox{$f_\mathrm{L,R}(E)=f(E-eV_\mathrm{L,R})$} are the shifted Fermi functions of the macroscopic reservoirs into which semi-infinite leads terminate; $V_\mathrm{b}=V_\mathrm{L} - V_\mathrm{R}$ is the applied bias voltage between them; ${\bm \Sigma}_\mathrm{L,R}(E,V_\mathrm{L,R})$ are the self-energies of semi-infinite leads whose band bottom is shifted by the applied voltages; and $\boldsymbol{\Gamma}_\mathrm{L,R}(E) = i[\boldsymbol{\Sigma}_\mathrm{L,R}(E)- \boldsymbol{\Sigma}^{\dagger}_\mathrm{L,R}(E)]$ are the level broadening matrices. Due to nonorthogonality of the basis of localized orbitals $|\phi_i \rangle$, we also use the overlap matrix ${\bm \Lambda}$ composed of elements $\langle \phi_i|\phi_j\rangle$.   For lateral heterostructure in Fig.~\ref{fig:fig1}(b), all matrices depend on $k_y$, while for vertical heterostructure  in Fig.~\ref{fig:fig1}(a) they depend on $(k_x,k_y)$.

\begin{figure*}
	%\center
	\includegraphics[width=8.9cm,clip=true]{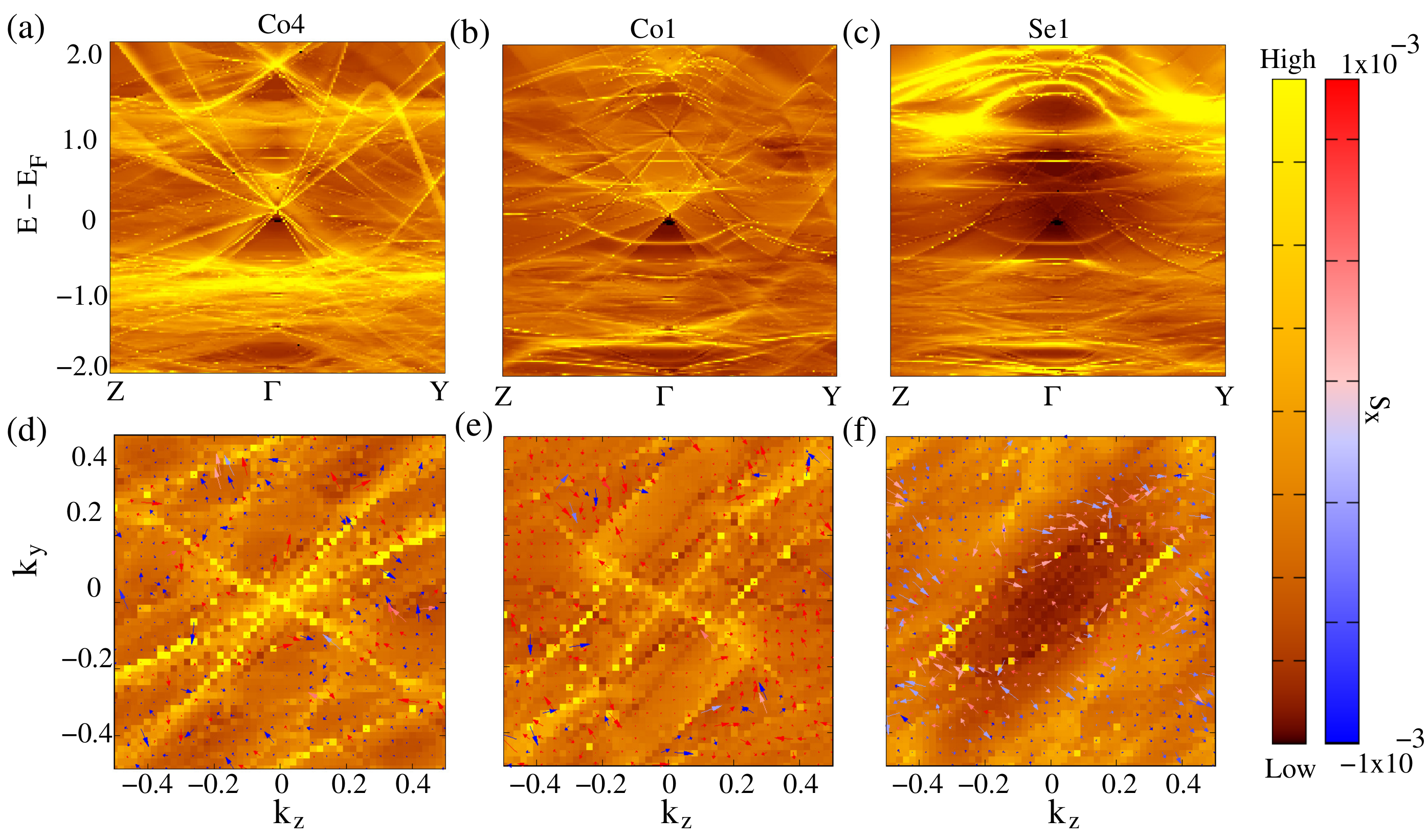}
	\includegraphics[width=8.9cm,clip=true]{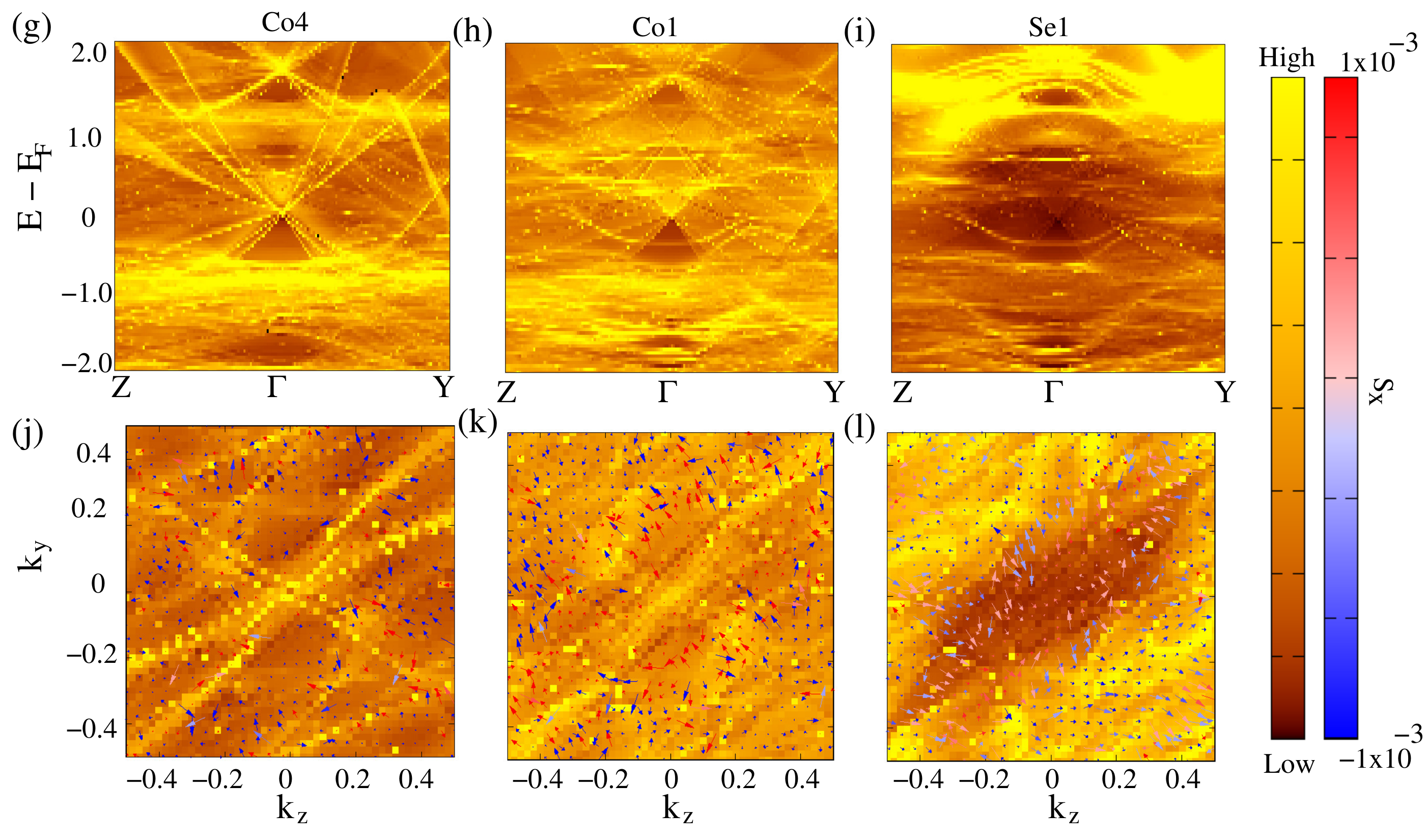}
	\includegraphics[width=8.9cm,clip=true]{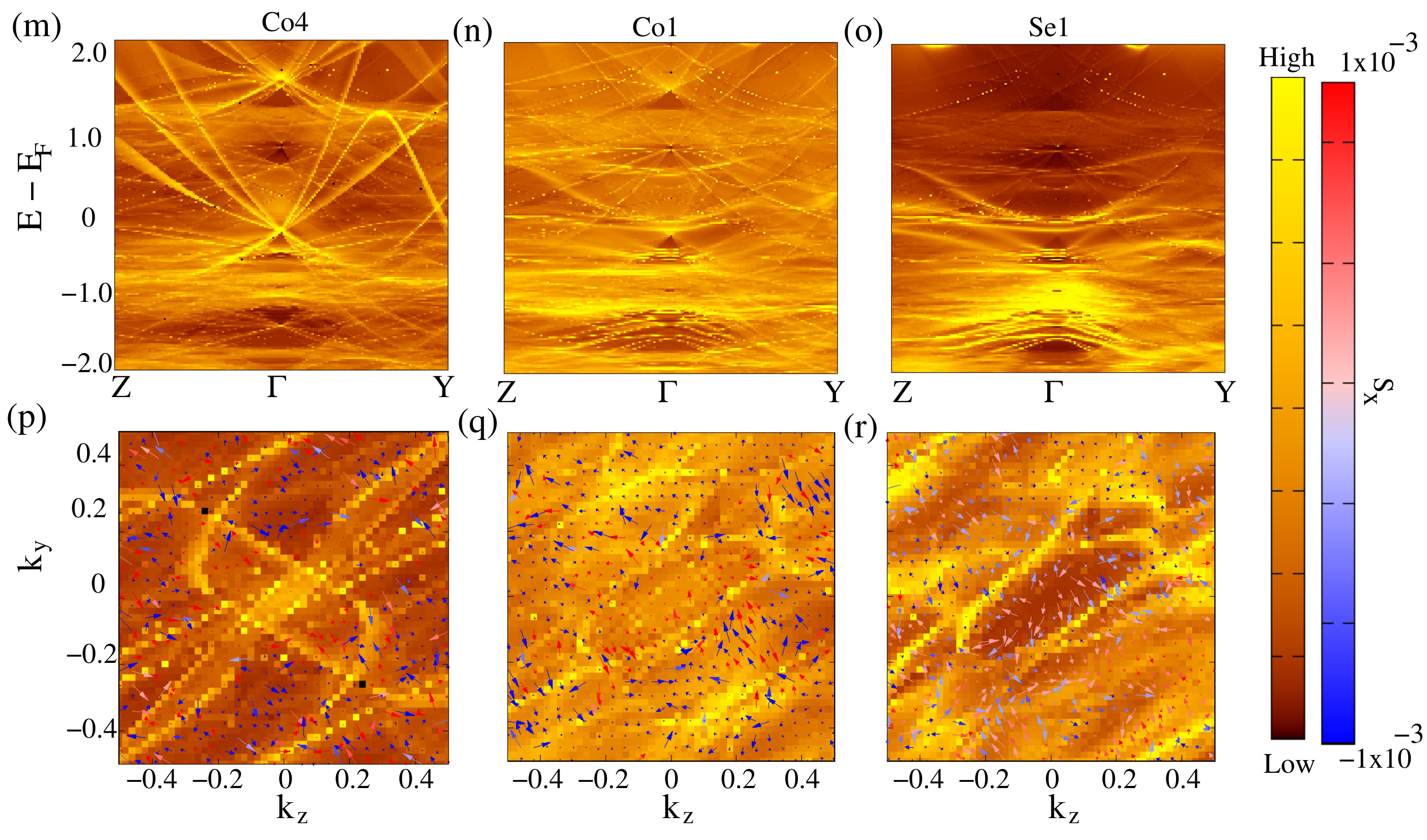}
	\includegraphics[width=8.9cm,clip=true]{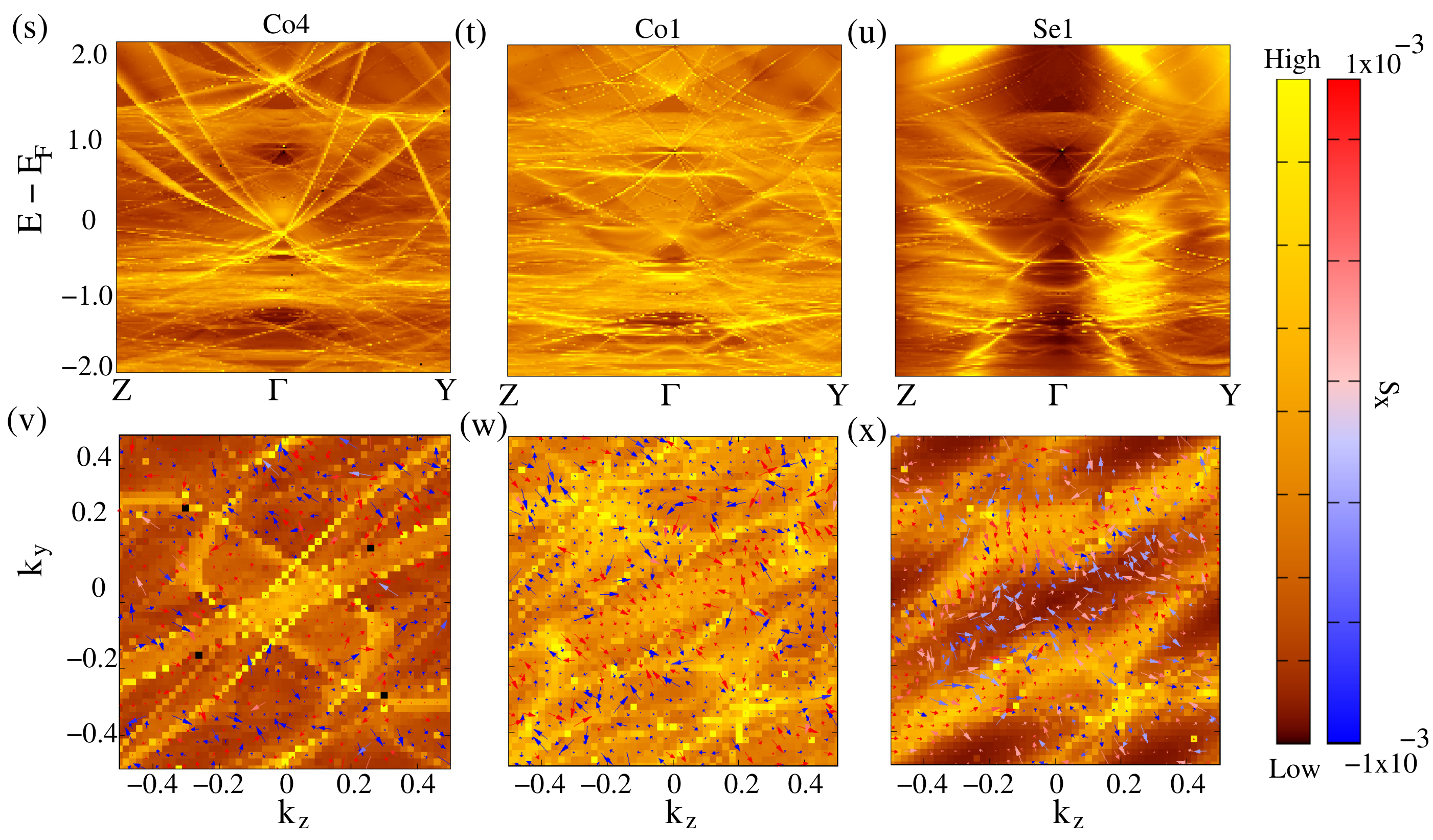}
	\caption{Spectral function $A(E; k_y,k_z, x\in\{{\rm Co4,Co1, Se1}\})$ plotted along the high-symmetry $k$-path, $Z-\Gamma-Y$ at the atomic plane passing through (a) Co4, (b) Co1 and (c) Se1  indicated in Fig.~\ref{fig:fig1}(a) for Co/1H-MoSe$_2$ heterostructure. The magnetization of semi-infinite Co layer is perpendicular to the interface, ${\bf m}{\rm_{Co}} \parallel \hat{x}$ [Fig.~\ref{fig:fig1}(a)]. Panels (d)--(f) plot constant energy contours of $A(E-E_F = 0; k_y,k_z, x \in \{{\rm Co4,Co1,Se1}\})$ at the Fermi energy $E-E_F=0$ and the corresponding spin textures within the monolayers Co4, Co1 and Se1, respectively, where the out-of-plane $S_x$ component of spin is indicated in color (red for positive and blue for negative). Panels (g)--(l), (m)--(r) and (s)--(x) show counterpart information to the respective panels (a)-(f) but for Co/1H-WSe$_2$, Co/1H-TaSe$_2$ and Co/1T-TaSe$_2$ heterostructure, respectively. The units for $k_y$ and $k_z$ are 2$\pi/a$ and 2$\pi/b$ where $a$ and $b$ are the lattice constants of  unit-cell of the heterostructure.}
	\label{fig:fig3}
\end{figure*} 

The spin-transfer torque is usually split into two components, $\mathbf{T}=\mathbf{T}^\mathrm{FL}+\mathbf{T}^\mathrm{DL}$. The field-like (FL) component  is even under  time-reversal and it affects magnetization precession around the effective magnetic field. The damping-like (DL) component is odd under time-reversal and it either enhances the Gilbert damping by pushing magnetization toward the effective magnetic field or it competes with the Gilbert term as ``antidamping.'' Due to its more complex angular dependence~\cite{Garello2013}, SO torque should be split into components which are odd or even in the magnetization  of a single FM layer like $\mathbf{m}_\mathrm{Co}$ in Fig.~\ref{fig:fig1}
\begin{equation}\label{eq:sotsplit}
\bold{T}_\mathrm{CD} = \bold{T}^{\rm odd}_\mathrm{CD} + \bold{T}^{\rm even}_\mathrm{CD}.
\end{equation}
Thus, it is also advantageous to decompose \mbox{${\bm \rho}_{\rm CD}={\bm \rho}^\mathrm{odd}_\mathrm{CD} + {\bm \rho}^\mathrm{even}_\mathrm{CD}$} into the  contributions whose trace with the Pauli matrices can yield directly $\mathbf{S}^{\rm odd}_\mathrm{CD}$ and  $\mathbf{S}^{\rm even}_\mathrm{CD}$, respectively, and the corresponding SO torque components  via Eq.~\eqref{eq:sotrealspace}. This is achieved by using the following expressions~\cite{Mahfouzi2016,Nikolic2018}
\begin{widetext}
\begin{subequations}\label{eq:rhosplit}
\begin{eqnarray}
{\bm \rho}^\mathrm{odd}_\mathrm{neq} & = & \frac{1}{8\pi} \int\limits_{-\infty}^{+\infty}\!\! dE  \left[f_\mathrm{L}(E) - f_\mathrm{R}(E) \right] \left(\mathbf{G} {\bm \Gamma}_\mathrm{L} \mathbf{G}^\dagger - \mathbf{G}^\dagger {\bm \Gamma}_\mathrm{L} \mathbf{G} -  \mathbf{G} {\bm \Gamma}_R \mathbf{G}^\dagger + \mathbf{G}^\dagger {\bm \Gamma}_R \mathbf{G} \right), \label{eq:rhodd} \\
{\bm \rho}^\mathrm{even}_\mathrm{neq} & = & \frac{1}{8\pi} \int\limits_{-\infty}^{+\infty}\!\! dE  \left[ f_\mathrm{L}(E) - f_\mathrm{R}(E) \right] \left( \mathbf{G} {\bm \Gamma}_\mathrm{L} \mathbf{G}^\dagger + \mathbf{G}^\dagger {\bm \Gamma}_\mathrm{L} \mathbf{G} -  \mathbf{G} {\bm \Gamma}_R \mathbf{G}^\dagger - \mathbf{G}^\dagger {\bm \Gamma}_R \mathbf{G} \right)  + \frac{1}{2\pi} \int\limits_{-\infty}^{+\infty}\!\! dE [f_\mathrm{L}(E) + f_\mathrm{R}(E)] \mathrm{Im}\, \mathbf{G}  \label{eq:rhoeven} \nonumber \\ 
 && - \frac{1}{\pi} \int\limits_{-\infty}^{\infty} dE \mathrm{Im}\, \bold{G}_0(E) f(E). 
\end{eqnarray}
\end{subequations}
\end{widetext}
Here $\bold{G}_0(E)$ is retarded GF in equilibrium obtained from Eq.~\eqref{eq:retarded} by setting $V_\mathrm{L} = V_\mathrm{R} = 0$, and the last term in Eq.~\eqref{eq:rhoeven}  is subtracted equilibrium density matrix ${\bm \rho}_\mathrm{eq}$~\cite{Mahfouzi2013,Stefanucci2013}. In the limit of small bias voltage when compared to the Fermi energy $E_F$,  $eV_b \ll E_F$, we can replace $[f_\mathrm{L}(E) - f_\mathrm{R}(E)] \mapsto -\partial f/\partial E $. Then, in the spirit of the linear-response theory, one can use $\mathbf{G}(E) \mapsto \mathbf{G}_0(E)$ with no bias voltage and  thereby induced voltage drop across the active region in the Fermi surface terms in Eq.~\eqref{eq:rhosplit} which multiply $[f_\mathrm{L}(E) - f_\mathrm{R}(E)]$.  However, in the second term in Eq.~\eqref{eq:rhoeven} multiplying $[f_\mathrm{L}(E) + f_\mathrm{R}(E)]$ one needs to use the voltage drop~\cite{Mahfouzi2013,Belashchenko2019}. This also requires to assume presence of some type of disorder since voltage drop is not allowed in ballistic systems. To reduce the computational expense, one can assume linear potential drop and use equilibrium charge and spin densities for all atoms as inputs in the Kohn-Sham Hamiltonian [Eq.~\eqref{eq:hks}], instead of performing fully self-consistent calculations for the whole system~\cite{Belashchenko2019}.

\begin{figure*}
	%\center
	\includegraphics[scale=0.65]{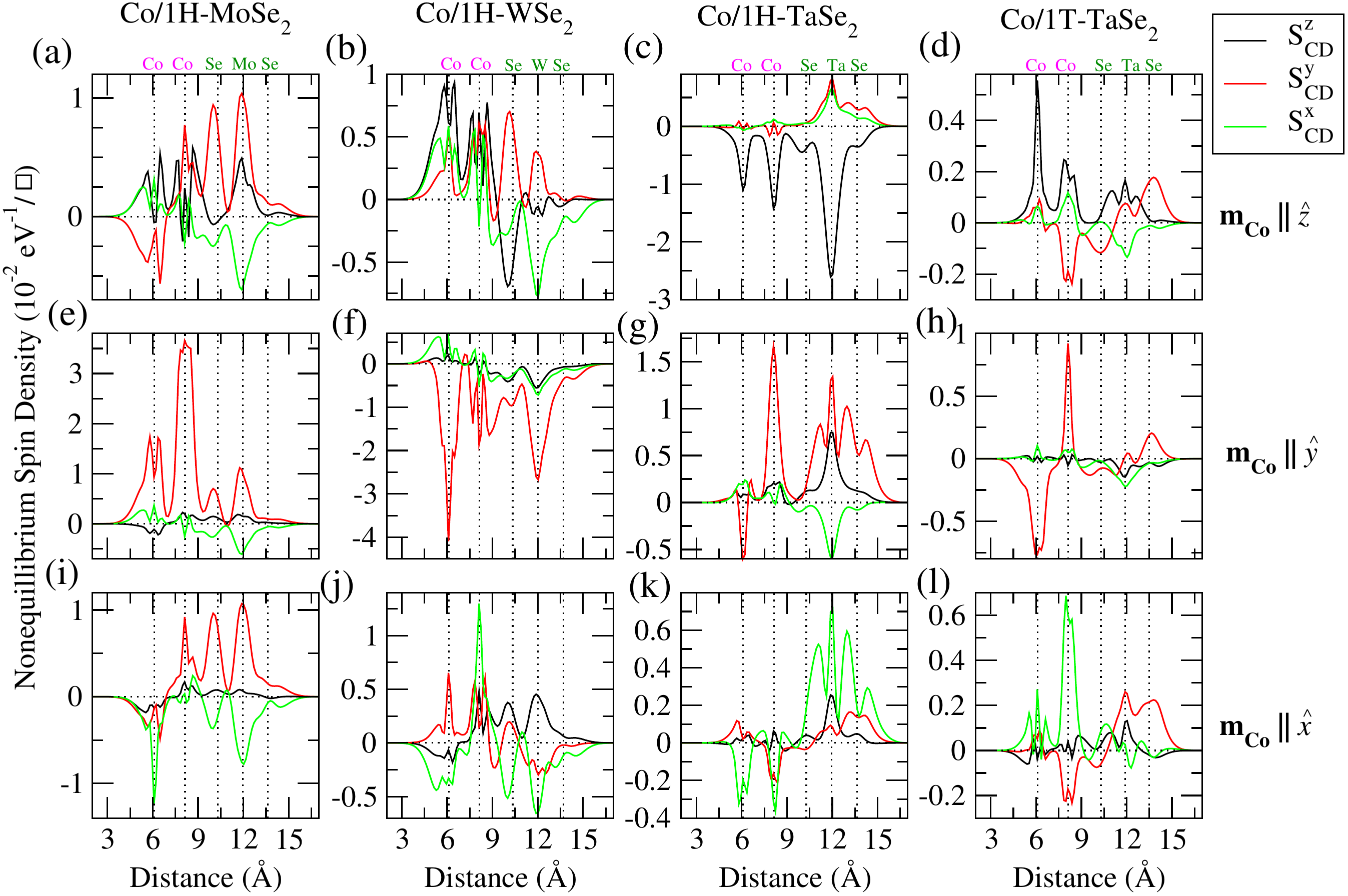}
	\caption{The spatial profile of nonequilibrium spin density vector $\mathbf{S}_{\rm CD}(z)$ in Eq.~\eqref{eq:scd} at the Fermi level for various Co/TMD heterostructures in Fig.~\ref{fig:fig1}(b) where TMD monolayer is: (a) 1H-MoSe$_2$; (b) 1H-WSe$_2$; (c) 1H-TaSe$_2$;  and (d) 1T-TaSe$_2$. The magnetization of Co layer  is perpendicular to the interface, ${\bf m}{\rm_{Co}}\parallel \hat{z}$. Black, red, and green lines indicate $z, y, {\rm and}~x$-components of $\mathbf{S}_{\rm CD}$, respectively. Panels (e)--(h) and (i)--(l) show the same informations as indicated, respectively, in the panels (a)--(d) for ${\bf m}{\rm_{Co}}\parallel \hat{y}$ and ${\bf m}{\rm_{Co}}\parallel \hat{x}$, respectively. Vertical dashed line indicates the position of each atom plane.}
	\label{fig:fig4}
\end{figure*}

Instead of real space calculations utilizing Eq.~\eqref{eq:sotrealspace}, for computational convenience we perform trace in the basis of localized pseudoatomic orbitals~\cite{Schlipf2015}. This yields nonequilibrium spin density as
\begin{equation}\label{eq:scd}
\mathbf{S}^\mathrm{even,odd}_{\rm CD}(k_y) = \mathrm{Tr}[\boldsymbol{\rho}^\mathrm{even,odd}_{\rm CD}(k_y)\boldsymbol{\sigma} \boldsymbol{\Lambda}^{-1}],
\end{equation}
and the corresponding components of SO torque
\begin{equation}\label{eq:sotvector}
\mathbf{T}^\mathrm{even,odd} = \frac {1}{\Omega_\mathrm{BZ}} \int_\mathrm{BZ}\! dk_y\, [\boldsymbol{\mathrm{S}}^\mathrm{even,odd}_{\rm CD}(k_y) \times \mathbf{B}_\mathrm{XC}(k_y)].
\end{equation}
An additional integration over the one-dimensional Brillouin zone (BZ) of length $\Omega_\mathrm{BZ}$ is performed because of assumed translational invariance of heterostructure in Fig.~\ref{fig:fig1}(b).

\subsection{Spectral functions and spin textures from equilibrium retarded GF}\label{sec:spectraltheory}
The spectral function (or local density of states) at an arbitrary plane at position $x$ within the active region is extracted from the retarded GF in equilibrium $\bold{G}_0(E)$ using
\begin{equation}\label{eq:spectral}
A(E;\mathbf{k}_{\parallel},x) = - \frac{1}{\pi}{\rm Im}[\bold{G}_0(E;\mathbf{k}_{\parallel};x,x)],
\end{equation}
where $\mathbf{k}_{\parallel}=(k_y,k_z)$ for the coordinate system in Fig.~\ref{fig:fig1}. The diagonal matrix elements of $\bold{G}_0(E;\mathbf{k}_{\parallel};x,x)$ are obtained from Eq.~\eqref{eq:retarded} by transforming it from orbital to real-space representation. At the chosen energy $E$ and thereby defined constant energy contours, we compute the spin textures from the spin-resolved spectral function. We note that computed spectral function and spin textures are exactly the same quantity that is measured on surfaces by spin-angle-resolved photoemission spectroscopy (spin-ARPES)~\cite{Tamai2013,Shoman2015}. However, spin-ARPES experiments cannot probe spectral function and spin textures on buried planes that are too far below the surface because the penetration depth of low-energy photons is 2--4 nm.

\begin{figure*}
	%\center
	\includegraphics[width= \linewidth,clip=true]{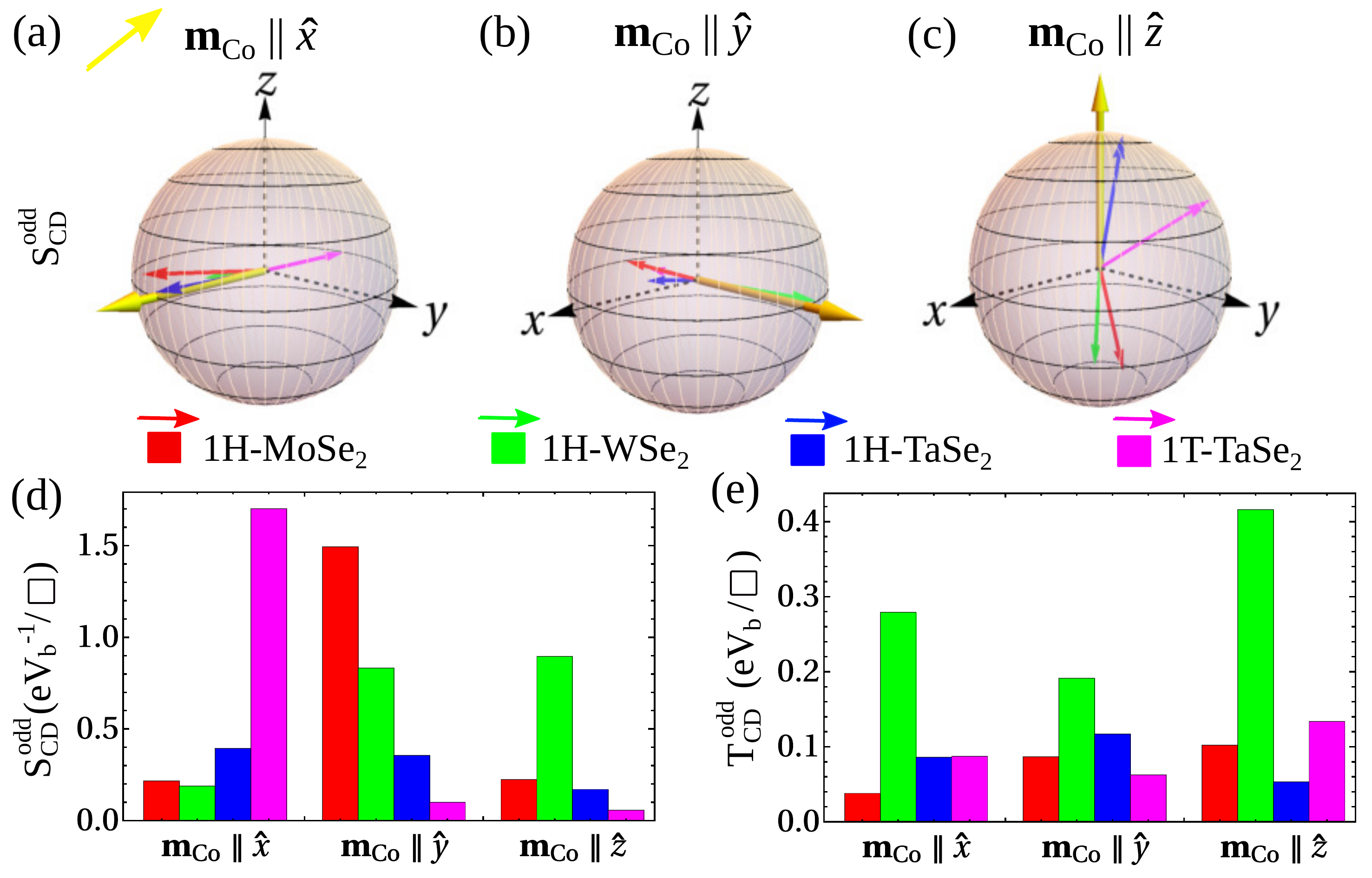}
	\caption{(a),(b),(c) Direction of current-driven nonequilibrium spin density vector $\mathbf{S}_{\rm CD}^{\rm odd}$ for magnetization $\mathbf{m}_{\rm Co}$ of Co layer in Fig.~\ref{fig:fig1}(b) oriented along the $x$-, $y$-, and $z$-axis, respectively. Red, green, blue and magenta arrows indicate the direction of $\mathbf{S}_{\rm CD}^{\rm odd}$ in  Co/1H-MoSe$_2$, Co/1H-WSe$_2$, Co/1H-TaSe$_2$ and Co/1T-TaSe$_2$ heterostructures, respectively. Thick yellow arrow indicates the direction of magnetization $\mathbf{m}_{\rm Co}$ of Co layer. Panels (d) and (e) are bar plots of magnitude of total $S_{\rm CD}^{\rm odd}=|\mathbf{S}_\mathrm{CD}|$ and $T_{\rm CD}^{\rm odd}=|T_{\rm CD}^{\rm odd}|$, respectively, within 2 MLs of Co layer embedded into   Co/1H-MoSe$_2$ (red), Co/1H-WSe$_2$ (green), Co/1H-TaSe$_2$ (blue), and Co/1T-TaSe$_2$ (magenta) heterostructures for different orientations of $\mathbf{m}_{\rm Co}$. The area of the common rectangular supercell of these heterostructures is denoted by $\Box = 4a\times~2\sqrt{3}a$, where $a$ is the lattice constant of the respective TMD.}
	\label{fig:fig5}
\end{figure*}

\subsection{Noncollinear density functional theory Hamiltonian of FM/TMD heterostructures}\label{sec:ncdft}

The ncDFT~\cite{Capelle2001,Eich2013a} operates with single-particle and spin-dependent Kohn-Sham (KS) Hamiltonian of heterostructures in Fig.~\ref{fig:fig1} which is given by 
\begin{eqnarray}\label{eq:hks}
\hat{H}_\mathrm{KS} & = & -\hbar^2\nabla^2/2m + \hat{V}_\mathrm{H}(\mathbf{r}) + \hat{V}_\mathrm{ext}(\mathbf{r}) + \hat{V}_\mathrm{XC}(\mathbf{r}) \nonumber \\  && + \hat{V}_\mathrm{SO}(\mathbf{r}) - {\bm \sigma} \cdot \mathbf{B}_\mathrm{XC}(\mathbf{r}). 
\end{eqnarray}
Here ${\hat{V}}_\mathrm{H}(\mathbf{r})$, ${\hat{V}}_\mathrm{ext}(\mathbf{r})$, and \mbox{${\hat{V}}_\mathrm{XC}(\mathbf{r}) = \delta E_\mathrm{XC}[n(\mathbf{r}),\mathbf{m}(\mathbf{r})]/\delta n(\mathbf{r})$} are  the Hartree potential, external potential and the XC potential, respectively; $\hat{V}_\mathrm{SO}$ is additional potential due to SO coupling; and the XC magnetic field, \mbox{$\mathbf{B}_\mathrm{XC}(\mathbf{r}) = \delta E_\mathrm{XC}[n(\mathbf{r}),\mathbf{m}(\mathbf{r})]/\delta \mathbf{m}(\mathbf{r})$}, is functional derivative 
of spin-dependent XC energy functional $E_\mathrm{XC}[n(\mathbf{r}),\mathbf{m}(\mathbf{r})]$ with respect to the vector  magnetization density $\mathbf{m}(\mathbf{r})$. The extension of DFT to the case of spin-polarized systems is formally derived in terms of $\mathbf{m}(\mathbf{r})$ and total electron density $n(\mathbf{r})$. In the collinear DFT $\mathbf{m}(\mathbf{r})$ points in the same direction at all points in space, while in ncDFT $\mathbf{m}(\mathbf{r})$ can point in an arbitrary direction~\cite{Capelle2001,Eich2013a}. The matrix representation of the XC magnetic field, employed in Eq.~\eqref{eq:sotvector}, can be extracted from the matrix representation $\mathbf{H}_\mathrm{KS}$ of $\hat{H}_\mathrm{KS}$ using   
\begin{equation}\label{eq:bxc}
\mathbf{B}_\mathrm{XC}=(2 \mathrm{Re}[\mathscr{H}^{\uparrow \downarrow}], -2\mathrm{Im}[\mathscr{H}^{\uparrow\downarrow}], \mathscr{H}^{\uparrow\uparrow} -  \mathscr{H}^{\downarrow \downarrow}),
\end{equation}
where $\mathscr{H} = \mathbf{H}_\mathrm{KS} - \mathbf{V}_\mathrm{SO}$.

We employ the interface builder in {\tt QuantumATK}~\cite{Smidstrup2019} package to construct a unit cell for the bilayer heterostructure and use the experimental lattice constants. The lattice strain between Co(0001) surface and monolayer TMDs is kept \mbox{$\lesssim 3$\%}. In order to determined the interlayer distance, we use Perdew-Burke-Ernzerhof (PBE) parametrization~\cite{Perdew1996} of the generalized gradient approximation (GGA) for the XC functional as implemented in {\tt QuantumATK} package~\cite{Smidstrup2019}. We find the average interlayer distance between Co and TMDs in Fig.~\ref{fig:fig1} in the range of $2.11-2.18$~\AA.

The matrix representation $\mathbf{H}_\mathrm{KS}$ employed in Eq.~\eqref{eq:retarded}, $\mathbf{H} \mapsto \mathbf{H}_\mathrm{KS}$, for the active region and self-energies ${\bm \Sigma}_\mathrm{L,R}(E)$ of semi-infinite leads for heterostructures in Fig.~\ref{fig:fig1} were computed using {\tt QuantumATK} where we employ PBE parametrization of GGA; norm-conserving fully relativistic pseudopotentials of the type SG15-SO~\cite{Schlipf2015} for describing electron-core interactions; and SG15 (medium) basis of localized pseudoatomic orbitals~\cite{Schlipf2015}. Periodic boundary conditions are employed in the plane perpendicular to the transport direction[which is along the $x$-axis in Fig.~\ref{fig:fig1}(b)], with a $1 \times 6$  $k$-point grid for self-consistent calculation. Nonequilibrium spin density and SO torque are then obtained by integrating over a denser  $1 \times 64$ $k$-point mesh. The energy mesh cutoff for the real-space grid is chosen as 100 Hartree. We use $15$~\AA~vacuum in the nonperiodic direction in order to remove the interaction between two periodic images.

\section{Results and Discussion}\label{sec:results}

\subsection{Atlas of spectral functions and spin textures in equilibrium}\label{sec:spectral}

Monolayer TMDs exhibit a wide-range of electronic properties---from semiconducting to semi-metallic and metallic---depending on the electronic occupation of transition metal atoms and its coordination with the chalcogen atoms. The coordination can be  either trigonal prismatic (1H) or octahedral phase (1T)~\cite{Zhang2016c}, as illustrated in Fig.~\ref{fig:fig1}(c). For example, 1H-MoSe$_2$ and 1H-WSe$_2$ are semiconductors with a gap of $\simeq 2$~eV, whereas both the 1H and 1T phases of TaSe$_2$ are metallic~\cite{Tran2016}. The 1H phase has space group  $D_{3h}^{1}$ (no. 187), which does not have the spatial inversion symmetry. On the other hand, 1T structure has space group of $D_{3d}^{1}$ (no. 164), which preserves the spatial inversion symmetry. Consequently, 1H phase of TMDs exhibits large SO splitting both in the conduction and valence bands~\cite{Zhu2011a}, while SO coupling does not have any significant effect on the bandstructure of 1T phase~\cite{Yan2015}. However, in the bilayer geometry in Fig.~\ref{fig:fig1}, the structural inversion symmetry is broken at the interface even for 1T-TMD due to the presence of Co layer.

The modification of spectral and spin textures within Co layer due to SO coupling ``injected'' by monolayer TMD and/or structural inversion asymmetry are shown in Fig.~\ref{fig:fig3}. For comparison, reference spectral functions and spin textures on the surface of semi-infinite Co in contact with vacuum can be found in Figs.~4(a)--(d) in Ref.~\cite{Marmolejo-Tejada2017}. Unlike those, where spin texture resemble the simple ones generated by the Rashba SO coupling in 2D electron gas with spin expectation values being tangent along the two Fermi circles, the  spin textures in Fig.~\ref{fig:fig3} are far more complex. Also, even though spectral functions on the Co side are similar when comparing Co/1H-MoSe$_2$ [Fig.~\ref{fig:fig3}(a)--(f)] with Co/1H-WSe$_2$ [Fig.~\ref{fig:fig3}(g)--(l)], or Co/1H-TaSe$_2$  [Fig.~\ref{fig:fig3}(m)--(r)] with  Co/1T-TaSe$_2$ [Fig.~\ref{fig:fig3}(s)--(x)], the corresponding spin textures at the constant energy contours of spectral functions can be quite different.  The appearance of  spin textures on the fourth ML of Co demonstrates how SO coupling from TMD propagates into the bulk of FM and away from the interface.

\begin{figure}
	%\center
	\includegraphics[width= \linewidth]{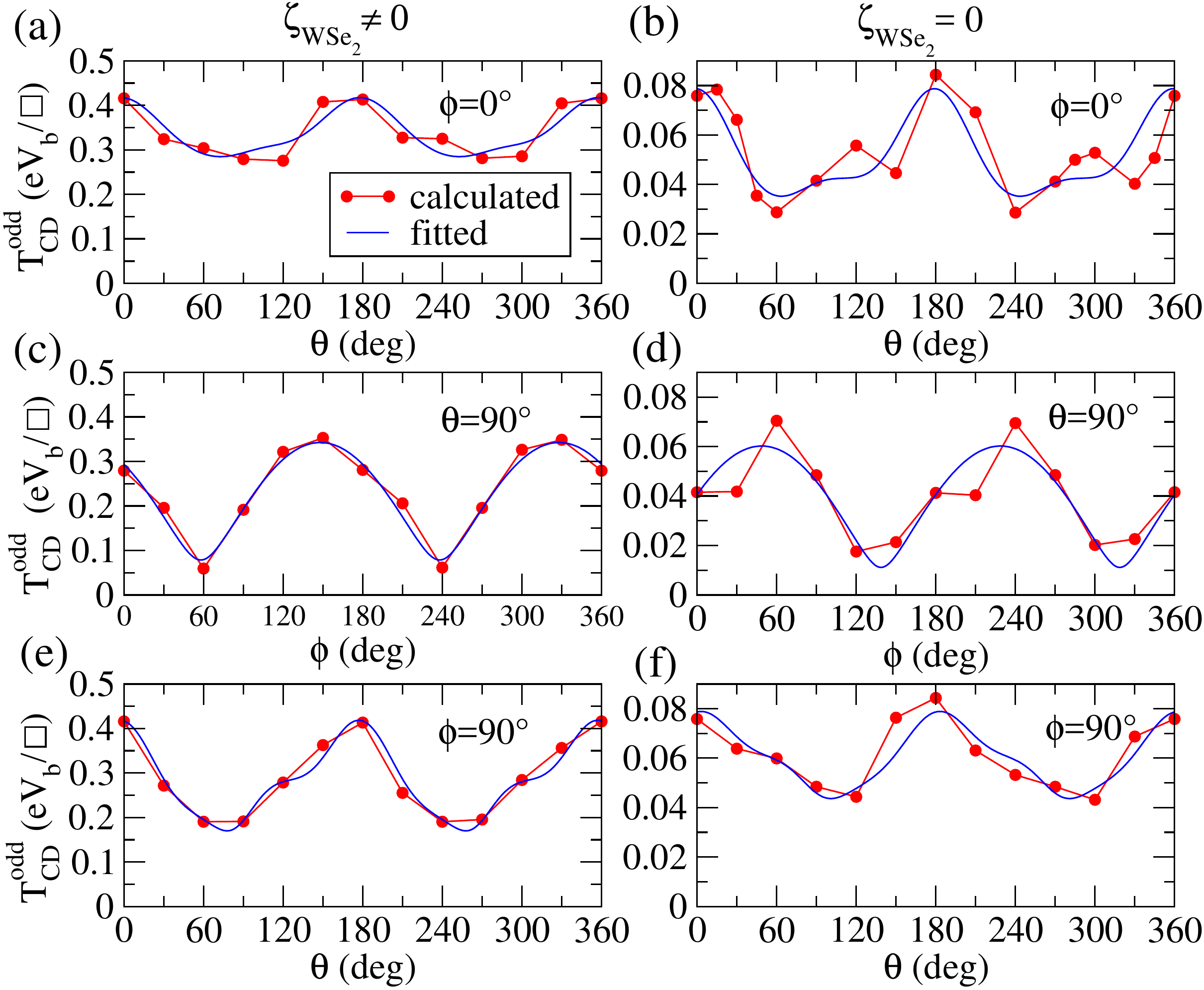}
	\caption{Azimuthal ($\theta$) or polar ($\phi$) angle dependence of the magnitude of odd component of SO torque, $T_{\rm CD}^{\rm odd} \equiv |\mathbf{T}_{\rm CD}^{\rm odd}|$ [Eq.~\eqref{eq:sotvector}], in Co/1H-WSe$_2$ heterostructure for: (a) $\phi=0^{\circ}$; (b) $\theta=90^{\circ}$; and (c) $\phi=90^{\circ}$. In the left (right) row of panels  $\mathbf{T}_{\rm CD}^{\rm odd}$ is computed with SO coupling in WSe$_2$ being turned on (off). Blue curves are fit to numerically computed red dots using expansion in Eq.~\eqref{eq:expansion} with nonzero coefficients provided in Table~\ref{tab:tab1}.}
	\label{fig:fig6}
\end{figure}

\subsection{Atlas of spin densities and SO torques out of equilibrium}\label{sec:sot}

When current flows through atomic planes with spin textures shown in Fig.~\ref{fig:fig3}, the imbalance between different $\mathbf{k}$ vectors is created which leads to nonzero nonequilibrium spin density $\mathbf{S}_\mathrm{CD}(z)$. Its spatial profile for four different Co/TMD heterostructures in shown in Fig.~\ref{fig:fig4} while orienting the magnetization of Co layer along three different axes. The integral  of $\mathbf{S}_\mathrm{CD}=\int  dz\, \mathbf{S}_\mathrm{CD}(z)$ within the volume of Co layer (composed of 2 MLs) gives total nonequilibrium spin density, and the orientation of its vector is visualized in Fig.~\ref{fig:fig5}(a)--(c).

\begin{figure}
	%\center
	\includegraphics[scale=0.6]{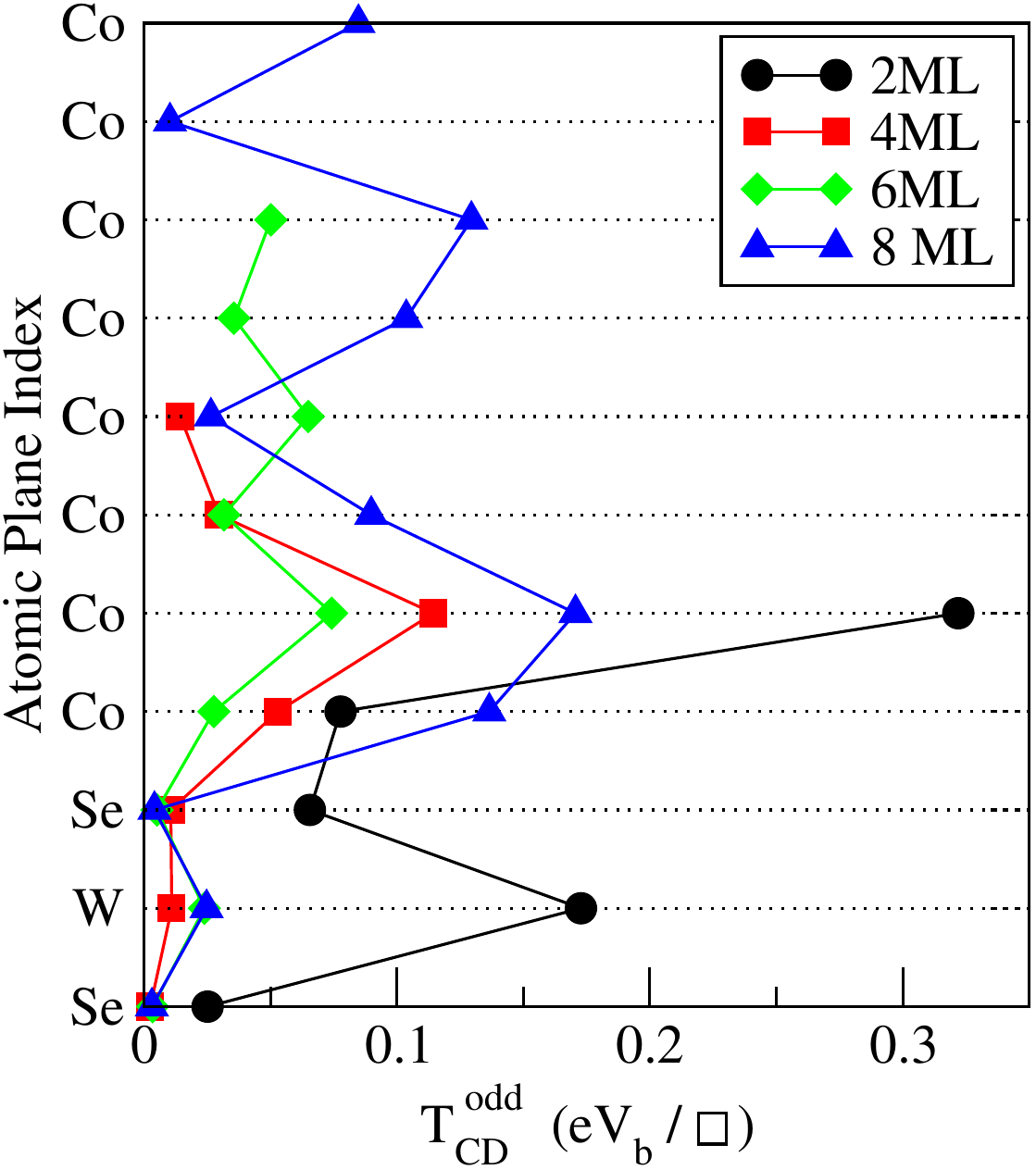}
	\caption{Layer resolved magnitude of odd component of SO torque $T_\mathrm{CD}^\mathrm{odd} \equiv |\mathbf{T}_\mathrm{CD}^\mathrm{odd}|$ in n-ML-Co/monolayer-1H-WSe$_2$ heterostructures where Co layer consists of 2 MLs (as in the illustration in Fig.~\ref{fig:fig1}) or 4, 6  and 8 MLs. The unpolarized charge current is injected parallel to the interface [along the $x$-axis in Fig.~\ref{fig:fig1}(b)] and it flows through each of n MLs of Co layer. The total SO torque on Co layer is the sum of $\mathbf{T}_\mathrm{CD}^\mathrm{odd}$ vectors on each of its MLs.}
	\label{fig:fig7}
\end{figure}

Due to assumed ballistic transport and absence of any spin Hall current along the $z$-axis from the ``bulk'' of TMD layer, only $\mathbf{S}_\mathrm{CD}^\mathrm{odd} \neq 0$ is generated in Fig.~\ref{fig:fig5}. The generation of $\mathbf{S}_\mathrm{CD}^\mathrm{even} \neq 0$ is also possible~\cite{Belashchenko2020}  in the presence of skew-scattering off impurities~\cite{Pesin2012a}, where the impurity themselves cab be  spin-independent~\cite{Zollner2019a,Sousa2020} but spin textures (like the ones in Fig.~\ref{fig:fig3}) contain an out-of-plane component~\cite{Milletari2017}. In the analysis of SO torque, based on intuition from  simplistic models like the 2D Rashba FM~\cite{Lee2015} or metallic surface of topological insulator covered by FM~\cite{Ndiaye2017}, one assumes that $\mathbf{S}_\mathrm{CD}^\mathrm{odd} || \hat{y}$ for current flowing along the $x$-axis, so that $\mathbf{T}_\mathrm{CD}^\mathrm{odd} \propto \hat{y} \times \mathbf{m}$. On the other hand, complexity of spin textures in Fig.~\ref{fig:fig3} leads to $\mathbf{S}_\mathrm{CD}^\mathrm{odd}$ which changes direction as we reorient $\mathbf{m}_\mathrm{Co}$. This leads to complex angular dependence of $\mathbf{T}_\mathrm{CD}^\mathrm{odd}$ component of SO torque, as often observed experimentally~\cite{Garello2013} and in first-principles quantum transport calculations~\cite{Belashchenko2019,Belashchenko2020,Dolui2020,Zollner2019a}.

Inasmuch as maximized value of $\mathbf{T}_\mathrm{CD}^\mathrm{odd}$ in Fig.~\ref{fig:fig5}(e) is found for Co/1H-WSe$_2$ heterostructure, we further  investigate such complex angular dependence of $\mathbf{T}_\mathrm{CD}^\mathrm{odd}$ for this bilayer in Fig.~\ref{fig:fig6}. For this purpose, we rotate   $\mathbf{m}_\mathrm{Co}$ within three different planes, and then fit the resulting SO torque curves using the following expansion~\cite{Garello2013,Belashchenko2019}
\begin{eqnarray}\label{eq:expansion}
\lefteqn{ \mathbf{T}^{\rm odd}_\mathrm{CD}  =  (\mathbf{p} \times \mathbf{m}_\mathrm{Co}) \left[ \sum_{n=0}^{\infty} T_{n\alpha} |\hat{z} \times \mathbf{m}_\mathrm{Co}|^{2n} \right] } \\ 
&+& \mathbf{m}_\mathrm{Co} \times (\mathbf{p} \times \mathbf{m}_\mathrm{Co})(\mathbf{m}_\mathrm{Co} \cdot \hat{x}) \left[ \sum_{n=0}^{\infty} T_{n\beta}| \hat{z} \times \mathbf{m}_\mathrm{Co}|^{2n} \right].
\end{eqnarray}
The unit vector $\mathbf{p}$ can be determined by symmetry arguments~\cite{Belashchenko2019}, or calculated where we give its direction in Table~\ref{tab:tab1} for both SO coupling switched on ($\zeta_{\rm WSe_2}\neq 0$) and off ($\zeta_{\rm WSe_2}=0$) in ncDFT calculations. The lowest order term  $T_{0\alpha}(\mathbf{p} \times \mathbf{m}_\mathrm{Co})$ in Eq.~\eqref{eq:expansion} is conventional  FL torque~\cite{Manchon2019,Lee2015}, while higher order  terms can have properties of both FL and DL torque~\cite{Belashchenko2020}. The value of $T_{0\alpha}$, together with other non-negligible coefficients in Eq.~\eqref{eq:expansion}, is given in Table~\ref{tab:tab1}. Interestingly, even if $\zeta_{\rm WSe_2}=0$, so that interfacial SO coupling  is possible only because of structural inversion asymmetry of the bilayer, these coefficients are nonzero in Table~\ref{tab:tab1}. Upon switching the SO coupling on, $\zeta_{\rm WSe_2}\neq 0$, the coefficients in Table~\ref{tab:tab1} increase about five times when compared to those computed with $\zeta_{\rm WSe_2}=0$.

\begin{table}[t!]
	\centering
	\begin{tabular}{c|c|c c c c c c}
		& $\mathbf{p}$ =($\theta$,$\phi$) & $T_{0\alpha}$ & $T_{1\alpha}$ & $T_{2\alpha}$ & $T_{3\alpha}$ & $T_{0\beta}$ & $T_{1\beta}$  \\
		\hline\hline
		$\zeta_{\rm WSe_2}\neq 0$ & (75$^{\circ}$,60$^{\circ}$) & 0.416 & -0.254 & 0.192 & 0.001 & -0.092 & 0.105 \\ \hline
		$\zeta_{\rm WSe_2}=0$     & (105$^{\circ}$,120$^{\circ}$) & 0.079 & -0.038 & 0.063 & -0.001 & 0.067 & -0.014 \\
		\hline
	\end{tabular}
	\caption{Coefficients (in units of ${\rm eV_b/\Box}$) in the angular dependence expansion of odd component of SO torque in Eq.~\eqref{eq:expansion} for Co/1H-WSe$_2$ heterostructure. The SO coupling is either switched off ($\zeta_{\rm WSe_2} = 0$) or on ($\zeta_{\rm WSe_2} \neq 0$) in ncDFT calculations. 
		The unit vector $\mathbf{p}$ in Eq.~\eqref{eq:expansion} is given in the second column for which $\mathbf{m}_\mathrm{Co} || \mathbf{p}$ leads to $\mathbf{T}^\mathrm{odd}_\mathrm{CD} \equiv 0$.}
	\label{tab:tab1}
\end{table}

Finally, Fig.~\ref{fig:fig7} demonstrates how SO proximity effect penetrating over increasing number of  Co MLs analyzed in Fig.~\ref{fig:fig3} manifests in the physics of SO torque. That is, each ML of Co which acquires spin textures in Fig.~\ref{fig:fig3} due to  SO coupling ``injected'' by monolayer TMD will also generate nonequilibrium spin density $\mathbf{S}_\mathrm{CD}$ once the current flows through such ML. The noncollinearity between $\mathbf{S}_\mathrm{CD}$ and $\mathbf{B}_\mathrm{XC}(\mathbf{r})$ within such ML then leads to nonzero local SO torque in Eq.~\eqref{eq:sotrealspace}.  Figure~\ref{fig:fig7} demonstrates that even 8th ML away from Co/1H-WSe$_2$ interface can exhibit SO torque, which means that SO proximity effect and thereby induced spin textures from Fig.~\ref{fig:fig3} can propagate all the way to the opposite edge of thin layers of Co, at least in the clean limit.	

\section{Conclusions}\label{sec:conclusions}

The recent experiments~\cite{Sklenar2016,Shao2016,MacNeill2017,Guimaraes2018,Lv2018,Shi2019} on SO torque and spin-to-charge conversion~\cite{Cheng2015} utilizing FM/TMD heterostructures have chosen combination of these materials using usual ``trial-and-error'' procedure or generic symmetry 
arguments~\cite{MacNeill2017}. The application-geared arguments for the usage of such class of devices with monolayer TMDs are based on the 
fact that in traditional FM/heavy-metal SO torque devices~\cite{Manchon2019}, with many monolayers of heavy-metal, spin Hall torque is interfacial in nature and the bulk of heavy-metal is of little use but nonetheless it carries charge current and causes heat dissipation. Furthermore, the ratio between FL and DL components of SO torque, or spin-relaxation rate in spin-to-charge conversion experiments,  can be effectively controlled via electric field by applying the back-gate voltage~\cite{Lv2018} to FM/TMD bilayers. 

In this study, we demonstrate ncDFT calculations to obtain accurate Hamiltonian of FM/TMD interface in the representation of localized pseudoatomic orbitals. Such first-principles Hamiltonian is then combined with equilibrium GFs of nonperiodic heterostructures to compute spectral functions and spin textures at an arbitrary plane. In addition, first-principles GFs can be employed in quantum transport calculations to obtain nonequilibrium spin density and thereby driven SO torque. This makes it possible to precisely determined the most optimal materials combination for SO torque applications, where we find maximized SO torque in Co/WSe$_2$ bilayers among Co/1H-MoSe$_2$, Co/1H-WSe$_2$, Co/1H-TaSe$_2$ and Co/1T-TaSe$_2$ screened. 

We note that SO proximity effect in a single ML of conventional $3d$ transition-metal ferromagnets Fe, Co, Ni due to adjacent TMD~\cite{Polesya2016}, topological insulator~\cite{Hou2019} or heavy metal (such as Pt~\cite{Simon2018}) layers has been studied recently with the focus on its impact on the Gilbert damping enhancement~\cite{Hou2019} or introduction of Dzyaloshinskii-Moriya interaction giving rise to skyrmions~\cite{Polesya2016,Simon2018}. In this study, we demonstrate that such SO proximity effect is not confined to just a single ML of conventional room temperature ferromagnetic metals,  but it can propagate over many of their MLs. Furthermore, any ML near the interface with TMD which becomes SO-proximitized, when brought out of equilibrium by passing current through it will generate nonequilibrium spin density as the  key resource~\cite{Manchon2019,Soumyanarayanan2016,Han2018} for variety of spintronic effects and applications.

\section{Acknowledgments}
This work was supported by the U.S. Department of Energy (DOE) Grant No. DE-SC0016380. The supercomputing time was provided by XSEDE supported by the U.S. National Science Foundation (NSF) Grant No. ACI-1053575.

%********************references************************************************************************

%BibTeX
%Windows:
%\bibliographystyle{IEEEtran}
%\bibliography{C:/BIBTEX/qttg}

\begin{thebibliography}{100}
	
\bibitem{Kircher1968}
C.~J. Kircher, Superconducting proximity effect of Nb, Phys. Rev. {\bf 168}, 437 (1968).

\bibitem{Hauser1969}
J. J. Hauser, Magnetic proximity effect, Phys. Rev. {\bf 187}, 580 (1969). 

\bibitem{Altland2000}
A.~Altland, B.~D. Simons, and D.~T. Semchuk, Field theory of mesoscopic fluctuations insuperconductor-normal-metal systems, Adv.  Phys. {\bf 49}, 321 (2000).

\bibitem{Gueron1996}
S.~Gu\'eron, H.~Pothier, N.~O. Birge, D.~Esteve, and M.~H. Devoret, Superconducting proximity effect probed on a mesoscopic length scale, Phys. Rev. Lett. {\bf 77}, 3025 (1996). 

\bibitem{Zutic2019}
I.~\v{Z}uti\'{c}, A.~Matos-Abiague, B.~Scharf, H.~Dery, and K.~Belashchenko, Proximitized materials, Mater. Today {\bf 22}, 85 (2019). 

\bibitem{Freericks2001}
J.~K. {Freericks}, B.~K. {Nikoli{\'c}}, and P.~{Miller}, Tuning a Josephson junction through a quantum critical point, Phys. Rev. B  {\bf 64}, 054511 (2001).

\bibitem{Freericks2002}
J.~K. {Freericks}, B.~K. {Nikoli{\'c}}, and P.~{Miller}, Optimizing the speed of a Josephson junction with dynamical mean field theory, Int. J. Mod. Phys. B {\bf 16}, 531 (2002). 

\bibitem{Nikolic2002b}
B.~K. {Nikoli{\'c}}, J.~K. {Freericks}, and P.~{Miller}, Suppression of the ``quasiclassical'' proximity gap in sorrelated-metal-superconductor structures,  Phys. Rev. Lett. {\bf 88}, 077002 (2002).

\bibitem{Buzdin2005}
A.~I. Buzdin, Proximity effects in superconductor-ferromagnet heterostructures, Rev. Mod. Phys. {\bf 77}, 935 (2005).

\bibitem{Komatsu2012}
K.~Komatsu, C.~Li, S.~Autier-Laurent, H.~Bouchiat, and S.~Gu\'eron, Superconducting proximity effect in long superconductor/graphene/superconductor junctions: From specular Andreev reflection at zero field to the quantum Hall regime, Phys. Rev. B {\bf 86}, 115412 (2012). 

\bibitem{Li2018a}
Q.~Li {\em et al.}, Proximity-induced superconductivity with subgap anomaly in type II Weyl semi-metal WTe$_2$, Nano Lett. {\bf 18}, 7962 (2018). 

\bibitem{Sillanpaa2001}
M.~A. Sillanp\"{a}\"{a}, T.~T. Heikkil\"{a}, R.~K. Lindell, and P.~J. Hakonen, Inverse proximity effect in superconductors near ferromagnetic material, Europhys. Lett. (EPL) {\bf 56}, 590 (2001).
	
\bibitem{Lim2013}
W.~L. Lim, N.~Ebrahim-Zadeh, J.~C. Owens, H.~G.~E. Hentschel, and S.~Urazhdin, Temperature-dependent proximity magnetism in Pt, Appl. Phys. Lett. {\bf 102}, 162404 (2013).
	
\bibitem{Zhu2018}
L.~J. Zhu, D.~C. Ralph, and R.~A. Buhrman, Irrelevance of magnetic proximity effect to spin-orbit torques in heavy-metal/ferromagnet bilayers,  Phys. Rev. B  {\bf 98}, 134406 (2018)
	
\bibitem{Peterson2018}
T.~A. Peterson, A.~P. McFadden, C.~J. Palmstr\o{}m, and P.~A. Crowell, Influence of the magnetic proximity effect on spin-orbit torque efficiencies in ferromagnet/platinum bilayers, Phys. Rev. B {\bf 97}, 020403 (2018).
	
\bibitem{Belashchenko2019}
K.~D. Belashchenko, A.~A. Kovalev, and M.~van Schilfgaarde, First-principles calculation of spin-orbit torque in a Co/Pt bilayer, Phys. Rev. Materials {\bf 3}, 011401 (2019). 

\bibitem{Dolui2017}
K.~Dolui and B.~K. Nikoli\'{c}, Spin-memory loss due to spin-orbit coupling at ferromagnet/heavy-metal interfaces: {\em Ab initio} spin-density matrix approach, Phys. Rev. B {\bf 96}, 220403(R) (2017). 

\bibitem{Shoman2015}
T.~Shoman, A.~Takayama, T.~Sato, S.~Souma, T.~Takahashi, T.~Oguchi, K.~Segawa, and Y.~Ando, Topological proximity effect in a topological insulator hybrid, Nat. Commun. {\bf 6},  6547 (2015).

\bibitem{Spataru2014}
C.~D. Spataru and F.~L\'eonard, Fermi-level pinning, charge transfer, and relaxation of spin-momentum locking at metal contacts to topological insulators, Phys. Rev. B {\bf 90}, 085115 (2014).

\bibitem{Marmolejo-Tejada2017}
J.~M. Marmolejo-Tejada, P.-H. Chang, P.~Lazi\'{c}, S.~Smidstrup, D.~Stradi, K.~Stokbro, and B.~K. Nikoli\'{c}, Proximity band structure and spin textures on both sides of topological-insulator/ferromagnetic-metal interface and their charge transport probes, Nano Lett.  {\bf 17}, 5626 (2017).

\bibitem{Zhang2016}
J.~Zhang, J.~P. Velev, X.~Dang, and E.~Y. Tsymbal, Band structure and spin texture of Bi$_2$Se$_3$/$3d$ ferromagnetic metal interface, Phys. Rev. B {\bf 94}, 014435 (2016).

\bibitem{Hsu2017}
Y.-T. Hsu, K.~Park, and E.-A. Kim, Hybridization-induced interface states in a topological-insulator/ferromagnetic-metal heterostructure, Phys. Rev. B {\bf 96} 235433 (2017).

\bibitem{Bansil2016}
A.~Bansil, H.~Lin, and T.~Das, \textit{Colloquium}: Topological band theory, Rev. Mod. Phys. {\bf 88}, 021004 (2016).
	
\bibitem{Nagaosa2013}
N.~Nagaosa and Y.~Tokura, Topological properties and dynamics of magnetic skyrmions, Nat. Nanotech. {\bf 8}, 899 (2013).
	
\bibitem{Boulle2016}
O.~Boulle {\em et al.}, Room-temperature chiral magnetic skyrmions in ultrathin magnetic nanostructures, Nat. Nanotech.  {\bf 11},  449 (2016).
	
\bibitem{Woo2016}
S.~Woo {\em et al.}, Observation of room-temperature magnetic skyrmions and their current-driven dynamics in ultrathin metallic ferromagnets, Nat. Mater.  {\bf 15}, 501 (2016).
	
\bibitem{Soumyanarayanan2017}
A.~Soumyanarayanan {\em et al.}, Tunable room temperature magnetic skyrmions in Ir/Fe/Co/Pt multilayers, Nat. Mater. {\bf 16}, 898 (2017).
	
\bibitem{Soumyanarayanan2016}
A.~Soumyanarayanan, N.~Reyren, A.~Fert, and C.~Panagopoulos, Emergent phenomena induced by spin-orbit coupling at surfaces and interfaces, Nature {\bf 539}, 509 (2016).

\bibitem{Simon2018}
E. Simon, L. R\'{o}zsa, K. Palot\'{a}s, and L. Szunyogh, Magnetism of a Co monolayer on Pt(111) capped by overlayers of 5d elements: A spin-model study, Phys. Rev. B {\bf 97}, 134405 (2018).
	
\bibitem{Tamai2013}
A.~Tamai, W.~Meevasana, P.~D.~C. King, C.~W. Nicholson, A.~de~la Torre, E.~Rozbicki, and F.~Baumberger, Spin-orbit splitting of the Shockley surface state on Cu(111), Phys. Rev. B {\bf 87}, 075113 (2013). 
	
\bibitem{Gmitra2016}
M.~Gmitra, D.~Kochan, P.~H\"ogl, and J.~Fabian, Trivial and inverted Dirac bands and the emergence of quantum spin Hall states in graphene on transition-metal dichalcogenides, Phys. Rev. B {\bf 93}, 155104 (2016).
	
\bibitem{Zollner2019b}
K.~Zollner and J.~Fabian, Single and bilayer graphene on the topological insulator Bi$_2$Se$_3$: Electronic and spin-orbit properties from first principles, Phys. Rev. B {\bf 100}, 165141 (2019).
	
\bibitem{Frank2016}
T.~Frank, M.~Gmitra, and J.~Fabian, Theory of electronic and spin-orbit proximity effects in graphene on Cu(111),  Phys. Rev. B {\bf 93}, 155142, (2016).

\bibitem{Island2019}
J. O. Island {\em et al.}, Spin-orbit-driven band inversion in bilayer graphene by the van der Waals proximity effect, Nature {\bf 571}, 85 (2019).
	
\bibitem{Dahal2014}
A.~Dahal and M.~Batzill, Graphene-nickel interfaces: A review, Nanoscale {\bf 6},  2548 (2014). 
	
\bibitem{Lazic2016}
P.~Lazi\'{c}, K.~D. Belashchenko, and I.~\v{Z}uti\'{c}, Effective gating and tunable magnetic proximity effects in two-dimensional heterostructures, Phys. Rev. B {\bf 93}, 241401 (2016). 
	
\bibitem{Hallal2017}
A.~Hallal, F.~Ibrahim, H.~Yang, S.~Roche, and M.~Chshiev, Tailoring magnetic insulator proximity effects in graphene: First-principles calculations, 2D Mater.  {\bf 4}, 025074 (2017).
		
\bibitem{Dolui2020}
K.~Dolui, M.~D. Petrovi\'{c}, K.~Zollner, P.~Plech\'{a}\v{c}, J.~Fabian, and B.~K. Nikoli\'{c}, Proximity spin-orbit torque on a two-dimensional magnet within van der Waals heterostructure: Current-driven antiferromagnet-to-ferromagnet reversible nonequilibrium phase transition in bilayer CrI$_3$, Nano Lett. {\bf 20}, 2288 (2020).
		
\bibitem{Manchon2019}
A.~Manchon, J.~\ifmmode~\check{Z}\else \v{Z}\fi{}elezn\'y, I.~M. Miron, T.~Jungwirth, J.~Sinova, A.~Thiaville, K.~Garello, and P.~Gambardella, Current-induced spin-orbit torques in ferromagnetic and antiferromagnetic systems, Rev. Mod. Phys. {\bf 91}, 035004 (2019).

\bibitem{Ramaswamy2018}
R.~Ramaswamy, J.~M. Lee, K.~Cai, and H.~Yang, Recent advances in spin-orbit torques: Moving towards device applications, Appl. Phys. Rev. {\bf 5}, 031107 (2018).

\bibitem{Han2018}
W.~Han, Y.~Otani, and S.~Maekawa, Quantum materials for spin and charge conversion, npj Quantum Mater. {\bf 3}, 27 (2018).

\bibitem{Tserkovnyak2005}
Y.~Tserkovnyak, A.~Brataas, G.~E.~W. Bauer, and B.~I. Halperin, Nonlocal magnetization dynamics in ferromagnetic heterostructures, Rev. Mod. Phys. {\bf 77}, 1375 (2005).
		
\bibitem{Chen2009}
S.-H. Chen, C.-R. Chang, J.~Q. Xiao, and B.~K. Nikoli\'{c}, Spin and charge pumping in magnetic tunnel junctions with precessing magnetization: A nonequilibrium Green function approach, Phys. Rev. B {\bf 79}, 054424 (2009). 
		
\bibitem{Mahfouzi2012}
F.~Mahfouzi, J.~Fabian, N.~Nagaosa, and B.~K. Nikoli\'{c}, Charge pumping by magnetization dynamics in magnetic and semimagnetic tunnel junctions with interfacial rashba or bulk extrinsic spin-orbit coupling, Phys. Rev. B {\bf 85}, 054406 (2012).
		
\bibitem{Dolui2019}
K.~Dolui, U.~Bajpai, and B.~K. Nikoli\'{c}, Spin-mixing conductance of ferromagnet/topological-insulator and ferromagnet/heavy-metal heterostructure: A first-principles Floquet-nonequilibrium Green function approach, {\tt arXiv:1905.01299} (2019).
		
\bibitem{Sanchez2013}
J.~C.~R. S\'{a}nchez, L.~Vila, G.~Desfonds, S.~Gambarelli, J.~P. Attan\'{e}, J.~M.~D. Teresa, C.~Mag\'{e}n, and A.~Fert, Spin-to-charge conversion using Rashba coupling at the interface between non-magnetic materials, Nat. Commun. {\bf 4}, 2944 (2013).
		
\bibitem{Mahfouzi2014a}
F.~Mahfouzi, N.~Nagaosa, and B.~K. Nikoli\'{c}, Spin-to-charge conversion in lateral and vertical topological-insulator/ferromagnet heterostructures with microwave-driven precessing magnetization, Phys. Rev. B {\bf 90}, 115432 (2014).
		
\bibitem{Shen2014}
K.~Shen, G.~Vignale, and R.~Raimondi, Microscopic theory of the inverse Edelstein effect, Phys. Rev. Lett. {\bf 112}, 096601 (2014).
	
\bibitem{Manchon2008}
A.~Manchon and S.~Zhang, Theory of nonequilibrium intrinsic spin torque in a single nanomagnet, Phys. Rev. B {\bf 78}, 21240 (2008).
		
\bibitem{Lee2015}
K.-S. Lee, D.~Go, A.~Manchon, P.~M. Haney, M.~D. Stiles, H.-W. Lee, and K.-J. Lee, Angular dependence of spin-orbit spin-transfer torques,  Phys. Rev. B {\bf 91}, 144401 (2015). 
		
\bibitem{Ndiaye2017}
P.~B. Ndiaye, C.~A. Akosa, M.~H. Fischer, A.~Vaezi, E.-A. Kim, and A.~Manchon, Dirac spin-orbit torques and charge pumping at the surface of topological insulators, Phys. Rev. B  {\bf 96}, 014408 (2017). 
		
\bibitem{Kim2017}
K.-W. Kim, K.-J. Lee, J.~Sinova, H.-W. Lee, and M.~D. Stiles, Spin-orbit torques from interfacial spin-orbit coupling for various interfaces,  Phys. Rev. B  {\bf 96}, 104438 (2017).

\bibitem{Amin2018}
V. P. Amin, J. Zemen, and M. D. Stiles, Interface-generated spin currents, Phys. Rev. Lett. {\bf 121}, 136805 (2018).

\bibitem{Ghosh2018}
S.~Ghosh and A.~Manchon, Spin-orbit torque in a three-dimensional topological insulator--ferromagnet heterostructure: Crossover between bulk and surface transport, Phys. Rev. B {\bf 97}, 134402 (2018). 
		

\bibitem{Haastrup2018}
S.~Haastrup {\em et al.},  The computational 2D materials database: high-throughput modeling and discovery of atomically thin crystals, 2D Mater. {\bf 5}, 
042002 (2018). 
		
\bibitem{Zhou2019a}
J.~Zhou {\em et al.}, 2DMatPedia, an open computational database of two-dimensional materials from top-down and bottom-up approaches, Sci. Data {\bf 6}, 86 (2019).
		
\bibitem{Andersen2015}
K.~Andersen, S.~Latini, and K.~S. Thygesen, Dielectric genome of van der Waals heterostructures, Nano Lett. {\bf 15}, 4616 (2015).
		
\bibitem{Stefanucci2013}
G.~Stefanucci and R.~van Leeuwen, \emph{Nonequilibrium Many-Body Theory of Quantum Systems: A Modern Introduction} (Cambridge University Press, Cambridge, 2013).
		
\bibitem{Wang2008b}
S.~Wang, Y.~Xu, and K.~Xia, First-principles study of spin-transfer torques in layered systems with noncollinear magnetization, Phys. Rev. B {\bf 77}, 184430 (2008).
		
\bibitem{Ellis2017}
M.~O.~A. Ellis, M.~Stamenova, and S.~Sanvito, e, Phys. Rev. B {\bf 96}, 224410 (2017).
	   
\bibitem{Nikolic2018}
B.~K. Nikoli\'{c}, K.~Dolui, M.~Petrovi\'{c}, P.~Plech\'{a}\v{c}, T.~Markussen, and K.~Stokbro, First-principles quantum transport modeling of
spin-transfer and spin-orbit torques in magnetic multilayers, in {\em Handbook of Materials Modeling: Applications: Current and Emerging Materials}, edited by W. Andreoni and S. Yip (Springer, Cham, 2018); {\tt arXiv:1801.05793}.		
		
\bibitem{Belashchenko2020}
K.~D. Belashchenko, A.~A. Kovalev, and M.~van Schilfgaarde, Interfacial contributions to spin-orbit torque and magnetoresistance in ferromagnet/heavy-metal bilayers, Phys. Rev. B {\bf 101}, 020407 (2020).
		
\bibitem{Mahfouzi2018}
F.~Mahfouzi and N.~Kioussis, First-principles study of the angular dependence of the spin-orbit torque in Pt/Co and Pd/Co bilayers, Phys. Rev. B {\bf 97}, 224426 (2018).

\bibitem{Mahfouzi2020}
F.~Mahfouzi, R.~Mishra, P.-H. Chang, H.~Yang, and N.~Kioussis, Microscopic origin of spin-orbit torque in ferromagnetic heterostructures: A first-principles approach, Phys. Rev. B {\bf 101}, 060405 (2020). 


\bibitem{Capelle2001}
K.~Capelle, G.~Vignale, and B.~L. {Gy\"{o}rffy}, Spin currents and spin dynamics in time-dependent density-functional theory, Phys. Rev. Lett. {\bf 87}, 206403 (2001).

\bibitem{Eich2013a}
F.~G. Eich and E.~K.~U. Gross, Transverse spin-gradient functional for noncollinear spin-density-functional theory, Phys. Rev. Lett. {\bf 111}, 156401 (2013).
		
\bibitem{Freimuth2014}
F.~Freimuth, S.~Bl\"ugel, and Y.~Mokrousov, Spin-orbit torques in Co/Pt(111) and Mn/W(001) magnetic bilayers from first principles, Phys. Rev. B {\bf 90}, 174423 (2014).
		
\bibitem{Mahfouzi2013}
F.~Mahfouzi and B.~K. Nikoli\'{c}, How to construct the proper gauge-invariant density matrix in steady-state nonequilibrium: Applications to spin-transfer and spin-orbit torques, SPIN {\bf 3}, 1330002 (2013).

\bibitem{Garello2013}
K.~Garello, I.~M. Miron, C.~O. Avci, F.~Freimuth, Y.~Mokrousov, S.~Bl\"{u}gel, S.~Auffret, O.~Boulle, G.~Gaudin, and P.~Gambardella, Symmetry and magnitude of spin-orbit torques in ferromagnetic heterostructures, Nat. Nanotech. {\bf 8}, 587 (2013).		
	
\bibitem{Mahfouzi2016}
F.~Mahfouzi, B.~K. Nikoli\'{c}, and N.~Kioussis, Antidamping spin-orbit torque driven by spin-flip reflection mechanism on the surface of a topological insulator: A time-dependent nonequilibrium Green function approach, Phys. Rev. B {\bf 93}, 115419 (2016).

\bibitem{Schlipf2015}
M.~Schlipf and F.~Gygi, Optimization algorithm for the generation of ONCV pseudopotentials, Comp. Phys. Commun. {\bf 196}, 36 (2015).

\bibitem{Smidstrup2019}
S.~Smidstrup {\em et al.} QuantumATK: an integrated platform of electronic and atomic-scale modelling tools, J. Phys.: Condens. Matter  {\bf 32}, 015901 (2019).  

\bibitem{Perdew1996}
J.~P. Perdew, K.~Burke, and M.~Ernzerhof, Generalized gradient approximation made simple, Phys. Rev. Lett. {\bf 77}, 3865 (1996).
						
\bibitem{Zhang2016c}
C.~Zhang {\em et al.}, Systematic study of electronic structure and band alignment of monolayer transition metal dichalcogenides in van der waals heterostructures, 2D Mater. {\bf 4}, 015026 (2016).
			
\bibitem{Tran2016}
M.~D. Tran, J.-H. Kim, and Y.~H. Lee, Tailoring photoluminescence of monolayer transition metal dichalcogenides, Curr. Appl. Phys. {\bf 16}, 1159 (2016). 
			
\bibitem{Zhu2011a}
Z.~Y. Zhu, Y.~C. Cheng, and U.~Schwingenschl\"ogl, Giant spin-orbit-induced spin splitting in two-dimensional transition-metal dichalcogenide semiconductors, Phys. Rev. B {\bf 84}, 153402 (2011).
			
\bibitem{Yan2015}
J.-A. Yan, M.~A.~D. Cruz, B.~Cook, and K.~Varga, Structural, electronic and vibrational properties of few-layer 2H- and 1T-TaSe$_2$, Sci. Rep. {\bf 5}, 16646, (2015).
			
\bibitem{Pesin2012a}
D.~A. Pesin and A.~H. MacDonald, Quantum kinetic theory of current-induced torques in Rashba ferromagnets, Phys. Rev. B {\bf 86}, 014416 (2012).

\bibitem{Sousa2020}
F. J. Sousa, G. Tatara, A. Ferreira, Emergent spin-orbit torques in two-dimensional material/ferromagnet interfaces, {\tt arXiv:2005.09670} (2020).

\bibitem{Zollner2019a}
K.~Zollner, M.~D. Petrovi\'{c}, K.~Dolui, P.~Plech\'{a}\v{c}, B.~K. Nikoli\'{c}, and J.~Fabian, Purely interfacial and highly tunable by gate or disorder spin-orbit torque in graphene doubly proximitized by two-dimensional ferromagnet Cr$_2$Ge$_2$Te$_6$ and monolayer WS$_2$, {\tt arXiv:1910.08072} (2019).

\bibitem{Milletari2017}
M. Milletar\`{\i}, M. Offidani, A. Ferreira, and R. Raimondi, Covariant conservation laws and the spin Hall effect in Dirac-Rashba systems, Phys. Rev. Lett. {\bf 119}, 246801 (2017).
			
\bibitem{Sklenar2016}
J.~Sklenar, W.~Zhang, M.~B. Jungfleisch, W.~Jiang, H.~Saglam, J.~E. Pearson, J.~B. Ketterson, and A.~Hoffmann, Perspective: Interface generation of spin-orbit torques, J.  Appl. Phys. {\bf 120}, 180901 (2016).
			
\bibitem{Shao2016}
Q.~Shao, G.~Yu, Y.-W. Lan, Y.~Shi, M.-Y. Li, C.~Zheng, X.~Zhu, L.-J. Li, P.~K. Amiri, and K.~L. Wang, Strong Rashba-Edelstein effect-induced spin–orbit torques in monolayer transition metal dichalcogenide/ferromagnet bilayers, Nano Lett. {\bf 16}, 7514 (2016).
		
\bibitem{MacNeill2017}
D.~MacNeill, G.~M. Stiehl, M.~H.~D. Guimaraes, R.~A. Buhrman, J.~Park, and D.~C. Ralph, Control of spin-orbit torques through crystal symmetry in WTe$_2$/ferromagnet bilayers, Nat. Phys. {\bf 13}, 300 (2017).
			
\bibitem{Guimaraes2018}
M.~H.~D. Guimar\~{a}es, G.~M. Stiehl, D.~MacNeill, N.~D. Reynolds, and D.~C. Ralph, Spin–orbit torques in nbse$_2$/permalloy bilayers, Nano Lett. {\bf 18},  1311 (2018).
			
\bibitem{Lv2018}
W.~Lv, Z.~Jia, B.~Wang, Y.~Lu, X.~Luo, B.~Zhang, Z.~Zeng, and Z.~Liu, Electric-field control of spin–orbit torques in WS$_2$/permalloy bilayers, ACS Appl. Mater. Interfaces {\bf 10},  2843 (2018).
			
\bibitem{Shi2019}
S.~Shi {\em et al.}, All-electric magnetization switching and Dzyaloshinskii-Moriya interaction in WTe$_2$/ferromagnet heterostructures, Nat. Nanotech. {\bf 14}, 945 (2019).
			
\bibitem{Cheng2015}
C.~Cheng {\em et al.},  Direct observation of spin-to-charge conversion in MoS$_2$ monolayer with spin pumping, {\tt arXiv:1510.0345} (2015).

\bibitem{Polesya2016}
S.~Polesya, S.~Mankovsky, D.~K\"{o}dderitzsch, W.~Bensch, and H.~Ebert, Dzyaloshinskii-Moriya interactions and magnetic texture in Fe films deposited on transition-metal dichalcogenides, Phys. Status Solidi RRL {\bf 10}, 218 (2016).

\bibitem{Hou2019}
Y.~Hou and R.~Wu, Strongly enhanced Gilbert damping in $3d$ transition-metal ferromagnet monolayers in contact with the topological insulator Bi$_2$Se$_3$, Phys. Rev. Applied {\bf 11}, 054032 (2019). 
			


		
\end{thebibliography}

\end{document}